\documentclass[]{jfm}

\usepackage{graphicx}
\usepackage{epstopdf,epsfig}
\usepackage{newtxtext}
\usepackage{newtxmath}
\usepackage{natbib}
\usepackage{hyperref}
\hypersetup{
    colorlinks = true,
    urlcolor   = blue,
    citecolor  = black,
}

\newcommand{\RomanNumeralCaps}[1]

\usepackage{latexsym}
\usepackage{amsmath}
\usepackage{amssymb}
\usepackage{tabularx}
\usepackage{multirow}
\usepackage{subfigure}
\usepackage{calligra}
\usepackage{calrsfs}
\usepackage{mathrsfs}
\usepackage{extarrows}
\usepackage{booktabs}
\usepackage{bm}
\usepackage{color}

\title{A kind of Lagrangian chaotic property of the Arnold-Beltrami-Childress flow}

\author{Shijie Qin\aff{2}
\and Shijun Liao\aff{1,2,3}
\corresp{\email{sjliao@sjtu.edu.cn}}
}

\affiliation{\aff{1}State Key Laboratory of Ocean Engineering, Shanghai 200240, China
\aff{2} Center of Marine Numerical Experiment, School of Naval Architecture, Ocean and Civil Engineering, Shanghai Jiao Tong University, Shanghai 200240, China
\aff{3} School of Physics and Astronomy, Shanghai Jiao Tong University, Shanghai 200240, China
}

\begin{document}
\maketitle

\begin{abstract}
Three-dimensional steady-state Arnold-Beltrami-Childress (ABC) flow has a chaotic Lagrangian structure, and also satisfies the Navier-Stokes (NS) equations with an external force per unit mass.
It is well-known that, although trajectories of a chaotic system have sensitive dependence on initial conditions, i.e. the famous ``butterfly-effect'', their statistical properties are often insensitive to small disturbances. This kind of chaos (such as governed by the Lorenz equations) is called normal-chaos. However, a new concept, i.e. ultra-chaos, has been reported recently, whose statistics are unstable to tiny disturbances. Thus, ultra-chaos represents higher disorder than normal chaos.
In this paper, we illustrate that ultra-chaos widely exists in Lagrangian trajectories of fluid particles in steady-state ABC flow. Moreover, solving the NS equation when $Re=50$ with the ABC flow plus a very small disturbance as the initial condition, it is found that trajectories of nearly all fluid particles become ultra-chaotic when the transition from laminar to turbulence occurs. These numerical experiments and facts highly suggest that ultra-chaos should have a relationship with turbulence. 
This paper identifies differences between ultra-chaos and sensitivity of statistics to parameters. Possible relationships between ultra-chaos and the Poincar\'{e} section, ultra-chaos and ergodicity/non-ergodicity, etc., are discussed. The concept of ultra-chaos opens a new perspective of chaos, the Poincar\'{e} section, ergodicity/non-ergodicity, turbulence and their inter-relationships.
\end{abstract}

\hspace{-0.4cm}{\bf Keyword} ABC flow, Lagrangian chaos, stability of statistics

\section{Introduction}

The Arnold-Beltrami-Childress (ABC) flow
\begin{eqnarray}
\mathbf{u}_{ABC}(x,y,z)
&=&[A\sin(z)+C\cos(y)]\,\mathbf{e}_x
+\;[B\sin(x)+A\cos(z)]\,\mathbf{e}_y  \nonumber \\
&+& [C\sin(y)+B\cos(x)]\,\mathbf{e}_z    \label{ABC-0}
\end{eqnarray}
describes a kind of stationary flow of incompressible fluid with periodic boundary conditions, where $\mathbf{u}_{ABC}$ is the velocity vector field, $A$, $B$ and $C$ are arbitrary constants, $x$, $y$ and $z$ are Cartesian coordinates, $\mathbf{e}_x$, $\mathbf{e}_y$ and $\mathbf{e}_z$ are the direction vectors of the Cartesian coordinate system, respectively. The ABC flow was first discovered by \citet{arnold1965} as a class of steady-state solutions of the Euler equations or the Navier-Stokes (NS) equations with external force per unit mass, and since then the Lagrangian chaotic property \citep{dombre1986chaotic, galloway1986dynamo, galloway1987note} and the so-called Beltrami property, i.e. substantial helicity $\mathbf{u}_{ABC}\hspace{0.2mm}\times\hspace{0.2mm}(\nabla\hspace{0.2mm}\times\hspace{0.2mm}\mathbf{u}_{ABC})=0$, of this kind of flow have aroused wide interest in nonlinear dynamics, hydrodynamics, and magnetohydrodynamics.

The property of exponential deviation of a fluid particle (i.e. Lagrangian chaos) in the above-mentioned ABC flow is typical of chaotic dynamical systems \citep{dombre1986chaotic, blazevski2014hyperbolic, didov2018analysis, didov2018nonlinear} and essential for the development of turbulent flows \citep{dombre1986chaotic, galloway1987note, podvigina1994non}. This feature, in conjunction with substantial helicity, is essential for fast dynamo action (i.e. fast generation of magnetic field in conducting fluids) \citep{moffatt1985topological, galloway1986dynamo, finn1988chaotic} and for the origin of magnetic field of large astrophysical objects \citep{childress1970new}.

For a chaotic dynamical system \citep{Li1975Period, Parker1989Practical, Lorenz1993The, PeterSmith1998Explaining, sprott2010, van2013shil, lee2014wind, gao2018flow}, sensitivity dependence on initial conditions (SDIC) of a trajectory was first discovered by \citet{poincare1890probleme} and then rediscovered by \citet{lorenz1963deterministic} who proposed the popular name ``butterfly-effect''.   In essence,  the SDIC reveals the {\em trajectory instability} of chaos.  Moreover, \citet{lorenz1989computational,lorenz2006computational} further discovered that the trajectories of chaotic dynamical systems have sensitive dependence {\em not only} on initial conditions (SDIC) {\em but also} on numerical algorithms (SDNA), because numerical noise, arising from truncation error and round-off error, is unavoidable for all numerical algorithms. All of these phenomena are based on the exponential increase of noise (or small disturbances), especially for the long-duration numerical simulation of a chaotic dynamical system \citep{1971On, 2020Superfast}. 
Naturally, the non-replicability/unreliability of  chaotic trajectories has certainly  led to heated debate about the credibility of numerical simulations of chaotic systems, with \citet{Teixeira2007Time} reaching the pessimistic conclusion that ``for chaotic systems, numerical convergence cannot be guaranteed forever''.

In order to gain a reproducible/reliable numerical simulation of chaos, \citet{Liao2009} proposed a numerical strategy, namely ``Clean Numerical Simulation'' (CNS) \citep{Liao2013, Liao2014, Liao2017-CNS-review}, to greatly reduce the background numerical noise arising from truncation and round-off errors over a sufficiently long interval of time for statistical properties to be evaluated. In the frame of the CNS \citep{Liao2009, Liao2013, Liao2014, Liao2017-CNS-review, LIAO2014On, hu2020risks, qin2020influence}, spatial and temporal truncation errors are reduced to a required tiny level by means of a fine enough spatial discretization (such as the spatial Fourier expansion) and a high enough order of Taylor expansion in the temporal dimension, respectively.  In particular, by using a large enough number of significant digits to represent all physical and numerical variables/parameters in multiple-precision floating-point arithmetic \citep{oyanarte1990mp}, the round-off error can be reduced to below a required tiny level. Furthermore, an additional simulation with even smaller level of background numerical noise is performed so as to determine the so-called ``critical predictable time'' $T_c$ by comparing such two simulations, so that their numerical noise (caused by truncation and round-off errors) can be negligible, i.e. several orders of magnitude smaller than the ``true'' physical solution, and thus the computer-generated trajectorie of chaos is reproducible/reliable within the whole spatial domain throughout the time interval $t\in[0,T_c]$. In this way, the CNS can provide reproducible/reliable trajectories of chaotic dynamical systems in an interval of time $[0,T_{c}]$ that is long enough for the statistics to be evaluated properly.

The CNS provides a useful tool by which to obtain reproducible/reliable simulations of chaotic trajectory over a prescribed long time duration. To date, CNS has been successfully applied to solve many chaotic dynamical systems, such as the Lorenz equations \citep{Liao2009, LIAO2014On}, two-dimensional turbulent Rayleigh-B\'{e}nard convection \citep{lin2017origin}, chaotic motion of a disk in free fall \citep{xu2021accurate}, and certain spatiotemporal chaotic systems such as the complex Ginzburg-Landau equation \citep{hu2020risks}, the damped driven sine-Gordon equation \citep{qin2020influence}, and so on. Using CNS, more than 2000 new families of periodic orbits of Newton's \citep{newton1687} three-body problem have been discovered \citep{Li2017More, li2018over, li2019collisionless}, which were also reported twice in {\em New Scientist} \citep{NewScientist2017, NewScientist2018}. It should be noted that only three families of periodic orbits of the three-body problem had been reported in the 300 years after Newton first posed the problem. Recently, comparing the CNS results (as benchmark solutions) with those given by the DNS (direct numerical simulation), \citet{Qin2022JFM} provided rigorous evidence that numerical noise acting as tiny artificial stochastic disturbances has both quantitative and qualitative influences on sustained turbulence. The foregoing illustrates the novelty, great potential, and validity of CNS for chaotic dynamic systems.

Obviously, the numerical simulation of chaotic trajectory given by CNS can be considered as benchmark solution by which to investigate the influence of numerical noise on chaos. Using CNS, it has been found that, for certain chaotic dynamical systems, such as the Lorenz equations \citep{lorenz1963deterministic}, which has one positive Lyapunov exponent, and the so-called hyper-chaotic Rossler system \citep{Stankevich2020Chaos}, which has two positive Lyapunov exponents, their statistics always remain the same under small disturbances, i.e. stable, although their trajectories are rather sensitive to small disturbances, i.e. unstable. The behavior of such systems can be classified as normal-chaos \citep{Liao2022AAMM}.  However, the statistical properties (such as the probability density function) of some other forms of chaos are extremely sensitive to {\em tiny} noise/disturbances \citep{Liao2022AAMM}, i.e. unstable, which is called ultra-chaos \citep{Liao2022AAMM, Yang2023CFS}.

\begin{table}
\tabcolsep 0pt
\vspace*{-2pt}
\begin{center}
\begin{tabular}{lcc}
 Type of dynamic systems \hspace{1.0cm}  & \hspace{1.0cm} Trajectory \hspace{1.0cm}  & \hspace{1.0cm}  Statistics  \hspace{1.0cm}  \\ \\
 non-chaos	&	stable	&	stable  \\
 normal-chaos	&	unstable	&	stable  \\
 ultra-chaos	&	unstable	&	unstable \\
\end{tabular}
\end{center}
\caption{Stability of trajectory and statistics of different types of dynamic systems.}    \label{disorder}
\end{table}

Why do we need such a new classification and such a new concept of ultra-chaos mentioned above? It is well-known that numerical noise, say,  truncation and round-off error, is unavoidable in numerical simulations. Thus, due to the famous butterfly-effect \citep{lorenz1963deterministic},  numerical noise of computer-generated simulations of a chaotic system exponentially enlarges so that numerical simulations quickly become a mixture of the ``true'' physical solution $s$ and the ``false'' numerical noise $\varepsilon$, which are mostly at the {\em same} order.  Any statistics, which are calculated using such kind of mixture, are based on a {\em hypothesis} that the statistics are {\em stable} to numerical noise. In other words, the statistics based on this kind of mixture (i.e. $s+\varepsilon$) are the same as those based on the ``true'' physical solution (i.e. $s$), say,   
\begin{equation}
\big<  s  +  \varepsilon \big>  =  \big< s  \big> \label{hypothesis}
\end{equation}
{\em must} hold, where $\big< \big >$ is a statistical operator. Here, the numerical noise $\varepsilon$ is in fact equivalent to a kind of small disturbance. Unfortunately, there exists {\em no} theoretical proof of this hypothesis, even though it is widely utilized in many publications. Is the hypothesis (\ref{hypothesis}) always true for {\em all} chaotic systems? The answer is unfortunately negative, according to \cite{Liao2022AAMM}, who proposed the new concept ``ultra-chaos'' and classify chaos into normal-chaos and ultra-chaos, as listed in table~\ref{disorder} for   
the stability of trajectory and statistics of different types of dynamic systems. Such a classification of chaos is clear and  easy to implement in practice. Several examples of ultra-chaos have been found in different types of chaotic systems \citep{Liao2022AAMM, Yang2023CFS} and even in a Rayleigh-B\'{e}nard turbulent flow \citep{Qin2022JFM}.              

In this paper, we use the unstable Arnold-Beltrami-Childress (ABC) flow (in the Lagrangian viewpoint) as an example to illustrate that ultra-chaos indeed widely exists and is in a higher disorder than a normal-chaos.  
Besides, we point out the essential differences between ultra-chaos and high sensitivity of statistics on certain parameters, and discuss possible relationships between ultra-chaos and ergodicity/non-ergodicity, the Poincar\'{e} section, etc. Moreover, we numerically solve the Navier-Stokes equation using the ABC flow plus a small disturbance as the initial condition  so as to investigate the property of Lagrangian chaos of trajectories. Our results strongly suggest that turbulence should have a close relationship with ultra-chaotic trajectories, although the detailed mechanism is not yet fully understood, and thus warrants further study.

\section{Ultra-chaos in the ABC flows}

Let $x(t)$, $y(t)$ and $z(t)$ represent the location coordinates of a fluid particle, and $\dot{x}(t)$, $\dot{y}(t)$ and $\dot{z}(t)$ denote their temporal derivatives. Thus, in the Lagrangian sense, the motion of a fluid particle in ABC flow (\ref{ABC-0}) is governed by
\begin{equation}
\left\{
\begin{array}{l}
\dot{x}(t)=A\sin[z(t)]+C\cos[y(t)],       \\
\dot{y}(t)=B\sin[x(t)]+A\cos[z(t)],        \\
\dot{z}(t)=C\sin[y(t)]+B\cos[x(t)],
\end{array}
\right.  \label{ABC}
\end{equation}
 with the initial condition
\begin{equation}
(x(0), y(0), z(0) ) = {\bf r}_0,    \label{ABC-ini}
\end{equation}
where ${\bf r}_0$ denotes a starting point of the fluid particle. Equation (\ref{ABC}) describes a typical conservative (i.e. volume-preserving) dynamical system. Without loss of generality, let us consider the case of $A=1$ and different values of $B$ and $C$. It should be emphasized here that, by means of CNS, we invariably obtain a reproducible/reliable trajectory of the chaotic motion of a fluid particle of the ABC flow over a sufficiently long interval of time.
To investigate the influence of small disturbance on trajectory of the fluid particle in ABC flow (\ref{ABC-0}) starting from $\mathbf{r}_{0}= (x(0), y(0), z(0))$, we compare the trajectories of two close fluid particles of the ABC flow, starting from the initial positions ${\bf r}_0 $ and ${\bf r}_{0}'={\bf r}_{0}+\hspace{0.2mm}(0,0,1)\hspace{0.2mm}\times\hspace{0.2mm}\delta$, respectively, where $\delta = |{\bf r}_{0} - {\bf r}_{0}' |$ is a tiny constant. Note that $\delta = 0$ when ${\bf r}_{0} = {\bf r}_{0}'$, corresponding to non-disturbance.

For example, without loss of generality, let us consider the motion of a fluid particle of the ABC flow (in the Lagrangian sense) starting from the point ${\bf r}_{0} = (0,0,0)$ in the case for $A=1$ and different values of $B$ and $C$. In order to investigate its chaotic property, we compare the trajectory with that starting from a very close one ${\bf r}_{0}'={\bf r}_{0}+\hspace{0.2mm}(0,0,1)\hspace{0.2mm}\times\hspace{0.2mm}\delta$, where we choose either $\delta = 10^{-5}$ or $10^{-10}$. In each case, the chaotic simulation remains reproducible over the long interval $t\in[0,10000]$ by means of a parallel algorithm of the CNS using the $200$th-order Taylor expansion with the time step $\Delta t = 0.01$ and representing all data in $500$-digit multiple-precision  (MP) floating-point arithmetic, whose replicability/reliability is guaranteed via another CNS result with even smaller background numerical noise, given by the $205$th-order Taylor expansion (with the same time step) and $520$-digit multiple-precision floating-point arithmetic.

\begin{figure}
    \begin{center}
            \subfigure[]{\includegraphics[width=2.55in]{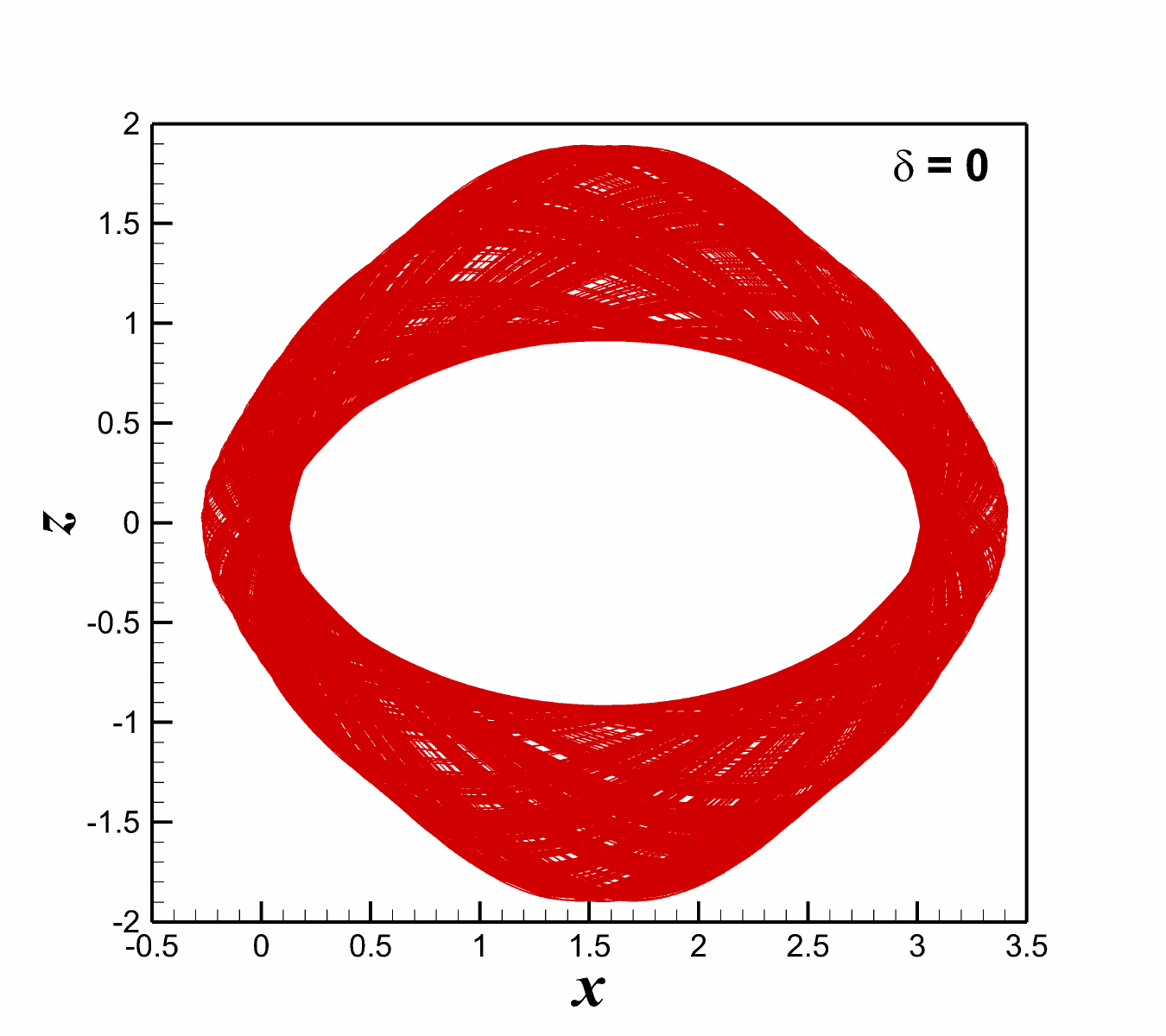}}
            \subfigure[]{\includegraphics[width=2.55in]{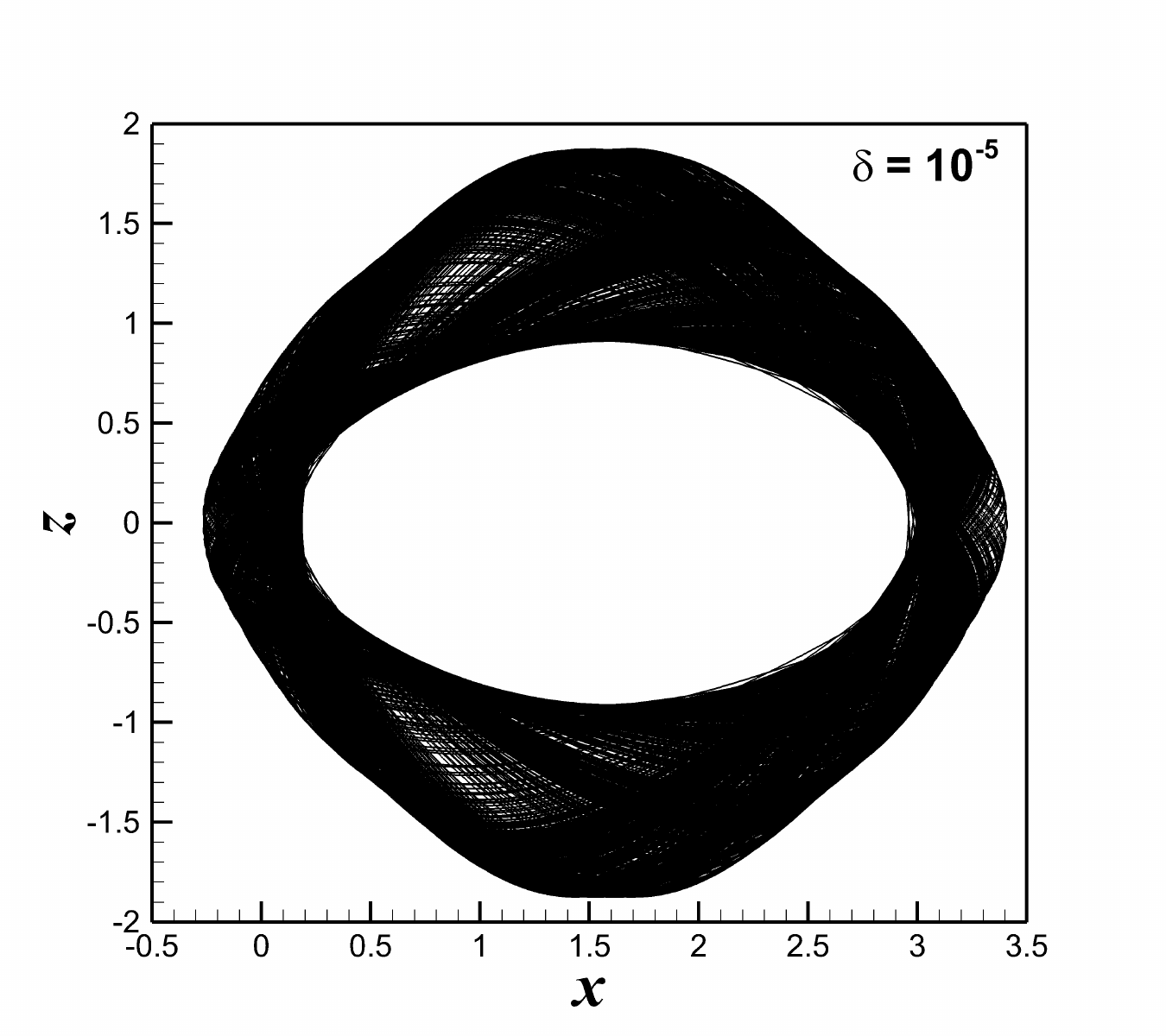}}\\
            \subfigure[]{\includegraphics[width=2.55in]{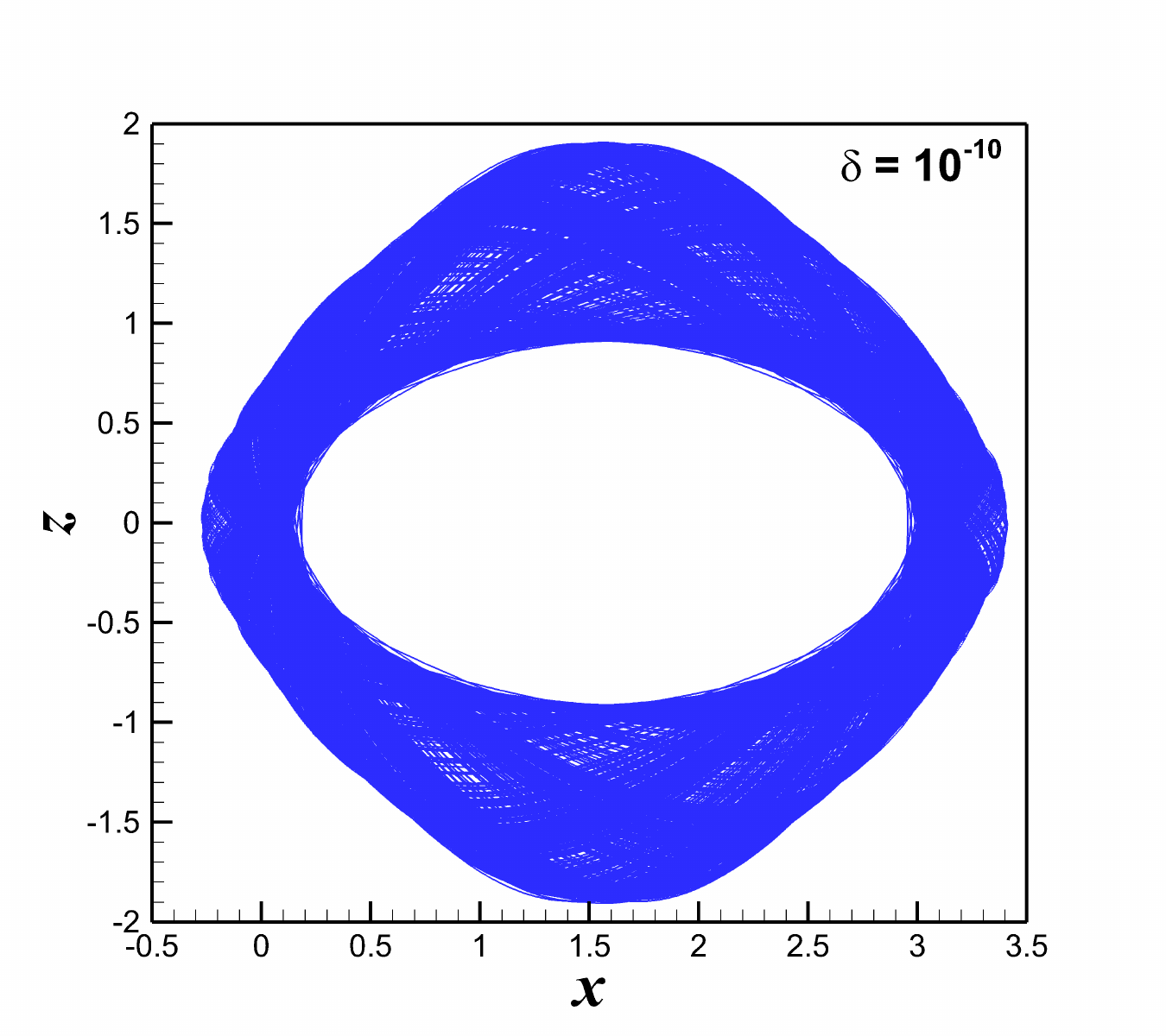}}
            \subfigure[]{\includegraphics[width=2.55in]{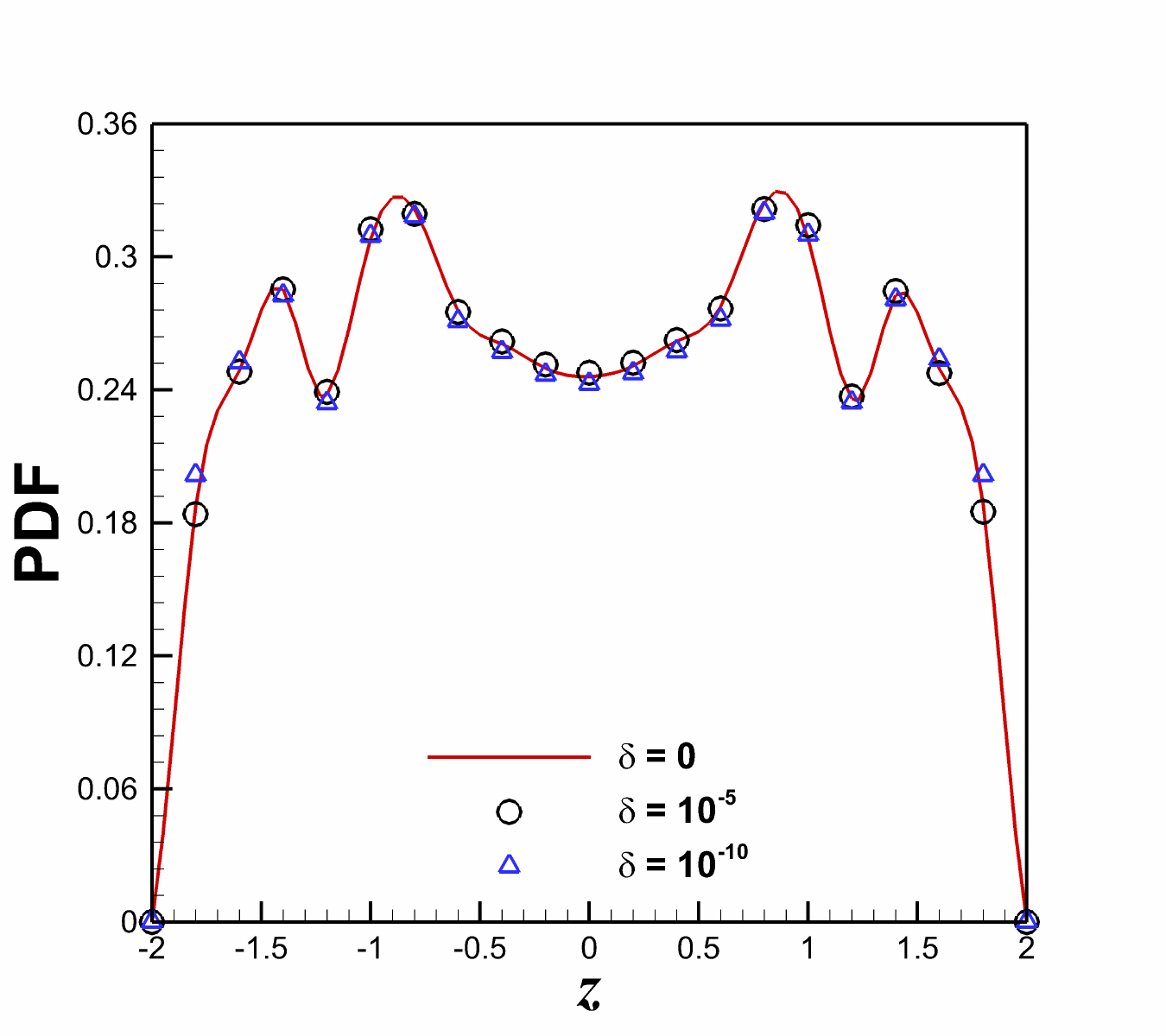}}
        \caption{Influence of tiny disturbances on the phase plot $x-z$ and the probability density function (PDF) of a normal-chaotic motion of a fluid particle in ABC flow. The curves are based on CNS results in $t\in[0,10000]$ of a normal-chaotic fluid particle, governed by ABC flow (\ref{ABC}) with (\ref{ABC-ini}) for $A=1$, $B=0.7$ and $C=0.42$ (with the maximum Lyapunov exponent $\lambda_{max}=0.01$), from the starting point ${\bf r}_{0}'=(0,0,0)+\hspace{0.2mm}(0,0,1) \times \delta$ when $\delta=0$ (red), $\delta=10^{-5}$ (black), and $\delta=10^{-10}$ (blue), respectively. (a) Phase plot $(x,z)$ when $\delta=0$; (b) Phase plot $(x,z)$ when $\delta=10^{-5}$; (c) Phase plot $(x,z)$ when $\delta=10^{-10}$; (d) PDFs of $z(t)$. }    \label{C042}
    \end{center}
\end{figure}

When $A=1$, $B=0.7$ and $C=0.42$, a fluid particle starting from ${\bf r}_{0}=(0,0,0)$ (corresponding to $\delta = 0$) experiences chaotic motion (with the maximum Lyapunov exponent $\lambda_{max}=0.01$) in a restricted spatial domain, as shown in figure~\ref{C042}(a) for its phase plot $(x,z)$ (considering the linear increase of value of $y$). For $\delta=10^{-5}$ and $\delta=10^{-10}$, although the chaotic trajectories of the two fluid particles, separately starting from the points  ${\bf r}'_{0}$ very close to ${\bf r}_{0}=(0,0,0)$, are rather sensitive to the starting point, their phase plots and statistical properties such as the probability density function (PDF) are almost the same as those given by the chaotic trajectory starting from ${\bf r}_{0}=(0,0,0)$ that corresponds to $\delta=0$, as shown in figure~\ref{C042}(b), (c), and (d), respectively. Note that here we show the PDFs (as well as other statistics) only of the $z$-coordinate values due to the similar properties of their $x$-coordinate counterparts. Therefore, for $A=1$, $B=0.7$ and $C=0.42$, the motion of the fluid particle starting from ${\bf r}_{0}=(0,0,0)$ is a normal-chaos, since its statistical properties such as the PDF of $z(t)$ are {\em stable}, i.e. not sensitive,  to a very small disturbance of the starting point.

\begin{figure}
    \begin{center}
            \subfigure[]{\includegraphics[width=2.55in]{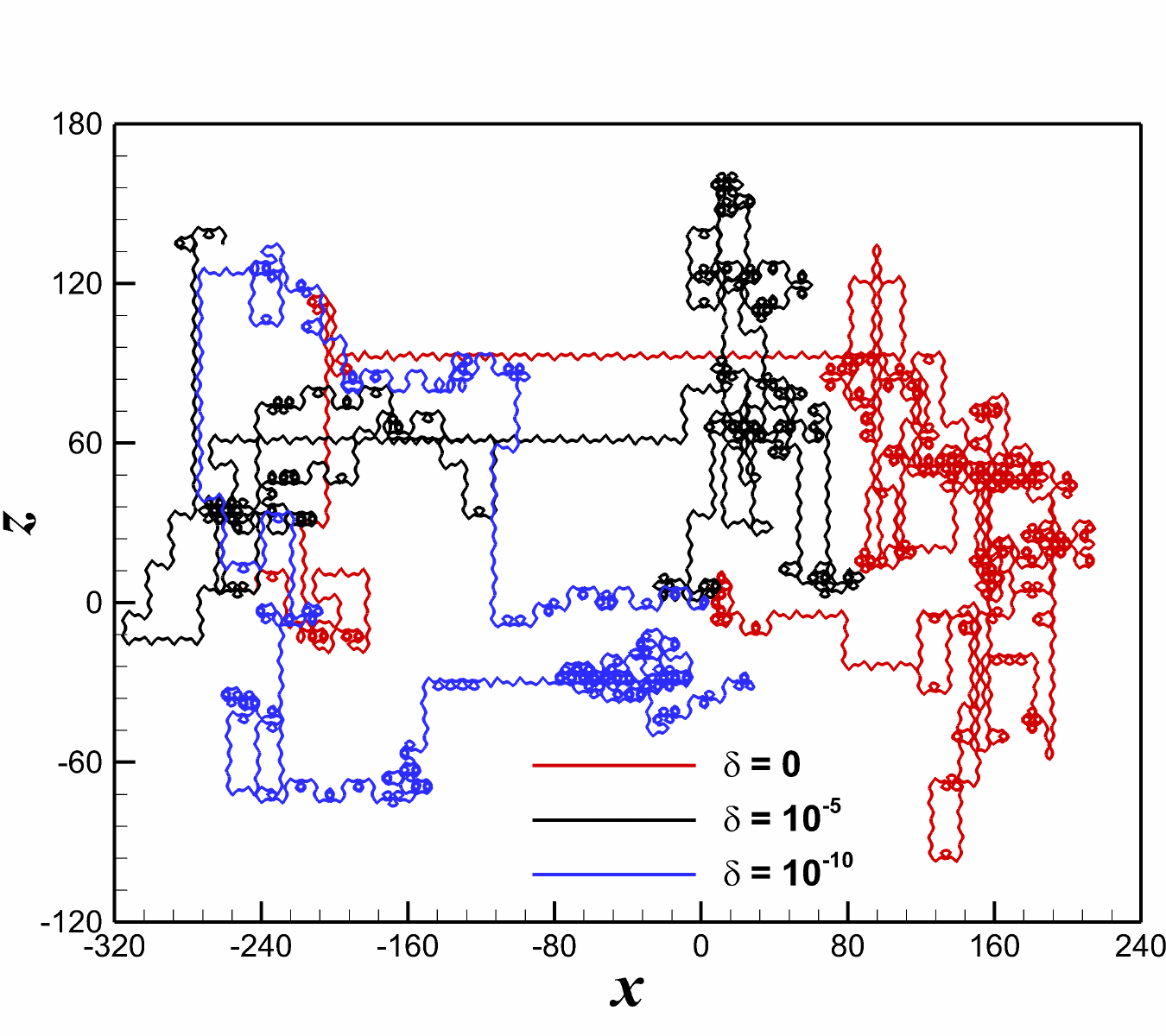}}
            \subfigure[]{\includegraphics[width=2.55in]{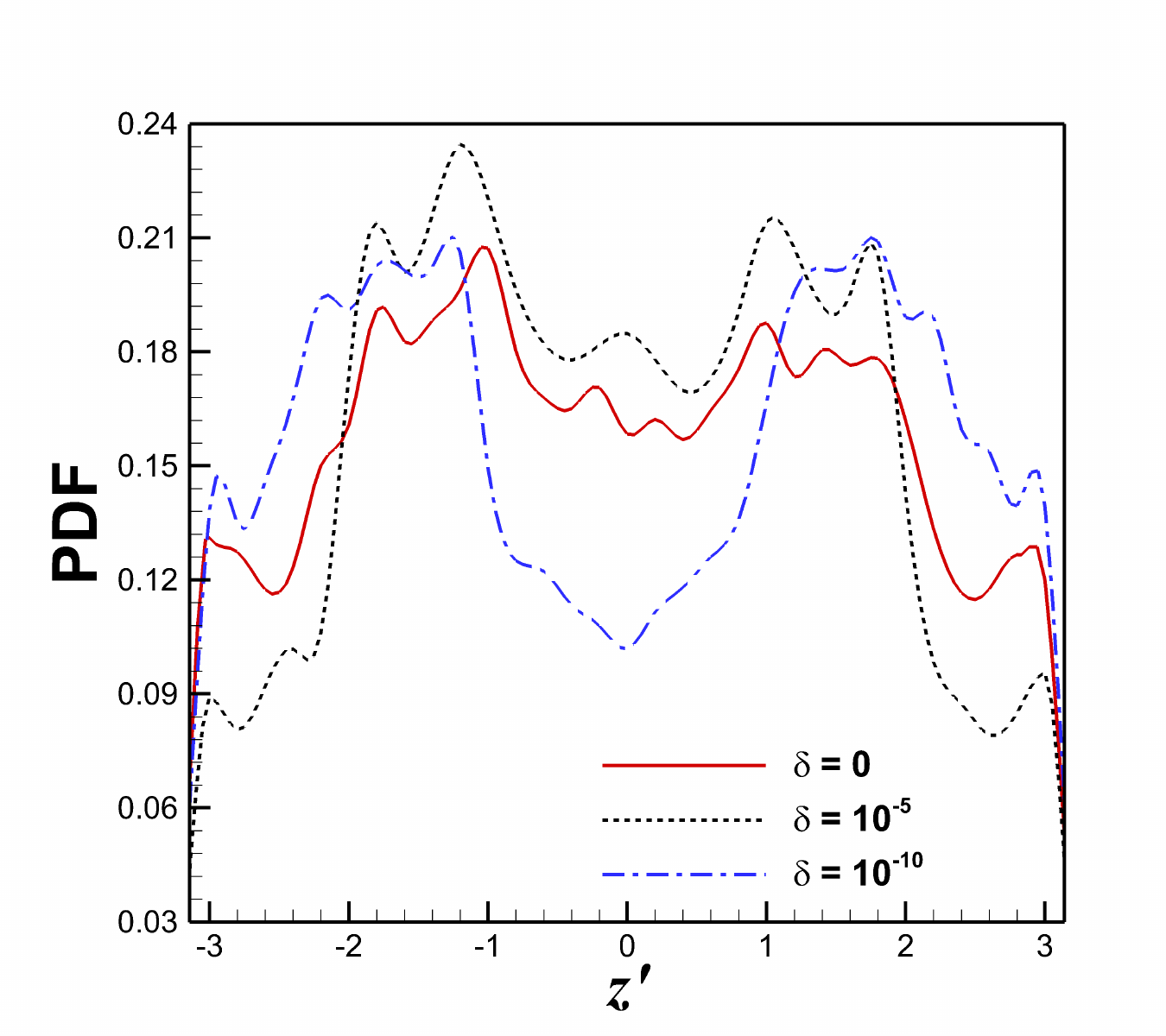}}
        \caption{Influence of tiny disturbances on the phase plot $x-z$ and the probability density function (PDF) of an ultra-chaotic motion of a fluid particle in ABC flow. The curves are based on CNS results in $t\in[0,10000]$ of an ultra-chaotic fluid particle, governed by ABC flow (\ref{ABC}) with (\ref{ABC-ini}) for $A=1.0$, $B=0.7$ and $C=0.43$ (with the maximum Lyapunov exponent $\lambda_{max}=0.06$) from the starting point ${\bf r}_{0}'=(0,0,0)+\hspace{0.2mm}(0,0,1) \times \delta $ when $\delta=0$ (red), $\delta=10^{-5}$ (black), and $\delta=10^{-10}$ (blue), respectively. (a) Phase plots $(x,z)$; (b) PDFs of the normalized results $z'(t)$. }    \label{C043}
    \end{center}
\end{figure}

However, for a small change in $C$, i.e.  $\Delta C = 0.01$, such that $A=1$, $B=0.7$ and $C=0.43$, the chaotic motion (with the maximum Lyapunov exponent $\lambda_{max}=0.06$) of the fluid particle of the ABC flow starting from ${\bf r}_{0} = (0,0,0)$ becomes quite different from that for $A=1$, $B=0.7$ and $C=0.42$:  the fluid particle moves further and further away from ${\bf r}_{0}$ and besides its phase plot $(x,z)$ becomes very sensitive to the small disturbance of the starting position, as shown in figure~\ref{C043}(a). These are quite different from the results in the case of $A=1$, $B=0.7$ and $C=0.42$. Since the ABC flow is periodic, we normalize the values of $z(t)$ to $[-\,\pi,\,\pi)$, i.e.
\begin{align}
z'(t)& = z(t)+2\pi\hspace{0.2mm} n_z,
\end{align}
where $n_z$ is integer, and  $-\,\pi \leq z' < +\,\pi $. Note that, as illustrated in figure~\ref{C043}(b), tiny disturbances in starting position can lead to huge deviations in the PDFs of the normalized chaotic simulations $z'(t)$ in $t\in[0,10000]$. In other words, for $A=1$, $B=0.7$ and $C=0.43$ in the ABC flow (\ref{ABC}), even statistical properties of the chaotic motion of the fluid particle starting from ${\bf r}_{0} = (0,0,0)$ are very sensitive to the initial position, and thus the corresponding motion of the particle is a kind of ultra-chaos.
Obviously, this kind of ultra-chaos is at a higher level of disorder than that of the normal-chaos, as shown in figure~\ref{C042} and figure~\ref{C043}. This example illustrates that ultra-chaos indeed exists  in the ABC flow.

\begin{table}
\tabcolsep 0pt
\vspace*{-2pt}
\begin{center}
\begin{tabular}{ccc}
 $\delta$ 	&	\hspace{1.0cm}  $\sigma^2$ 	\hspace{1.0cm}  & 	$\gamma_2$	\hspace{1.0cm}\\
\\
~~$0$ & $1.2$ & $1.8$~~ \\
~~$10^{-5}$ & $1.1$ & $1.8$~~ \\
~~$10^{-10}$ & $1.2$ & $1.8$~~ \\
\end{tabular}
\end{center}
\caption{Influence of tiny disturbances (i.e. $\delta$) on variance $\sigma^2$ and kurtosis $\gamma_2$ of the statistic results of $z(t)$ of the corresponding normal-chaotic trajectory of fluid particle in ABC flow for $A=1$, $B=0.7$ and $C=0.42$. These results are obtained by solving the chaotic dynamic system (\ref{ABC}) with (\ref{ABC-ini}) in $t\in[0,10000]$ by means of the CNS, from the starting point ${\bf r}_{0}'=(0,0,0)+ (0,0,1) \times \delta$ when $\delta=0$, $\delta=10^{-5}$, and $\delta=10^{-10}$, respectively.}    \label{moments-C042}
\end{table}

\begin{table}
\tabcolsep 0pt
\vspace*{-2pt}
\begin{center}
\begin{tabular}{ccc}
 $\delta$ 	&	\hspace{1.0cm}  $\sigma^2$ 	\hspace{1.0cm}  & 	$\gamma_2$	\hspace{1.0cm}\\
\\
$0$ 			& $2.9$ 	& 	$1.9$ 	\\
$10^{-5}$ 		& $2.5$ 	& 	$2.1$ 	\\
$10^{-10}$ 	& $3.4$ 	& 	$1.6$ 	\\
\end{tabular}
\end{center}
\caption{Influence of tiny disturbances (i.e. $\delta$) on variance $\sigma^2$ and kurtosis $\gamma_2$ of the statistic results of $z(t)$ of the corresponding ultra-chaotic trajectory of fluid particle in ABC flow for $A=1$, $B=0.7$ and $C=0.43$. These results are obtained by solving the chaotic dynamic system (\ref{ABC}) with (\ref{ABC-ini}) in $t\in[0,10000]$ by means of the CNS, from the starting point ${\bf r}_{0}'=(0,0,0)+ (0,0,1) \times \delta$ when $\delta=0$, $\delta=10^{-5}$, and $\delta=10^{-10}$, respectively.}    \label{moments-C043}
\end{table}

\begin{figure}
    \begin{center}
            \subfigure[]{\includegraphics[width=2.55in]{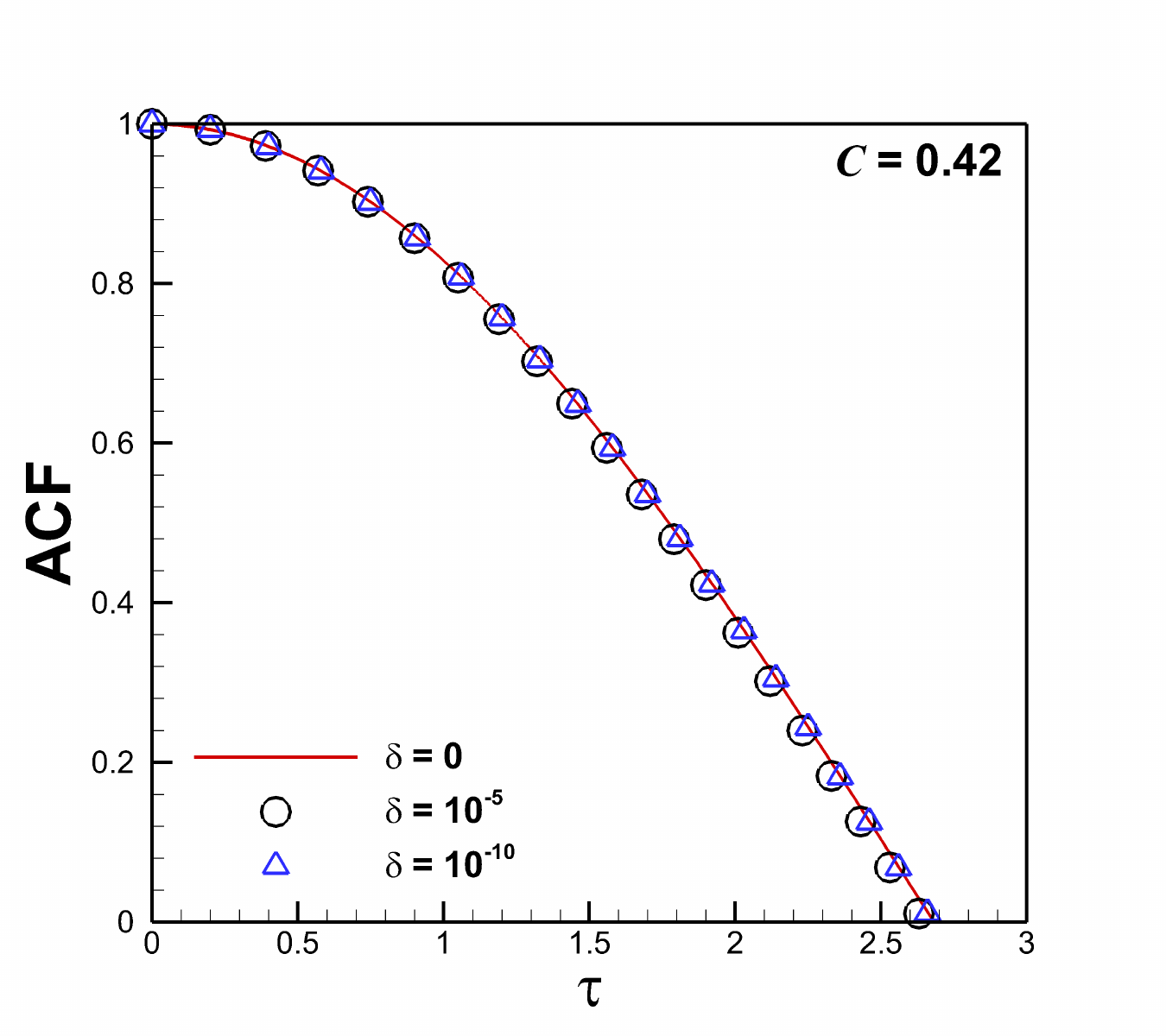}}
            \subfigure[]{\includegraphics[width=2.55in]{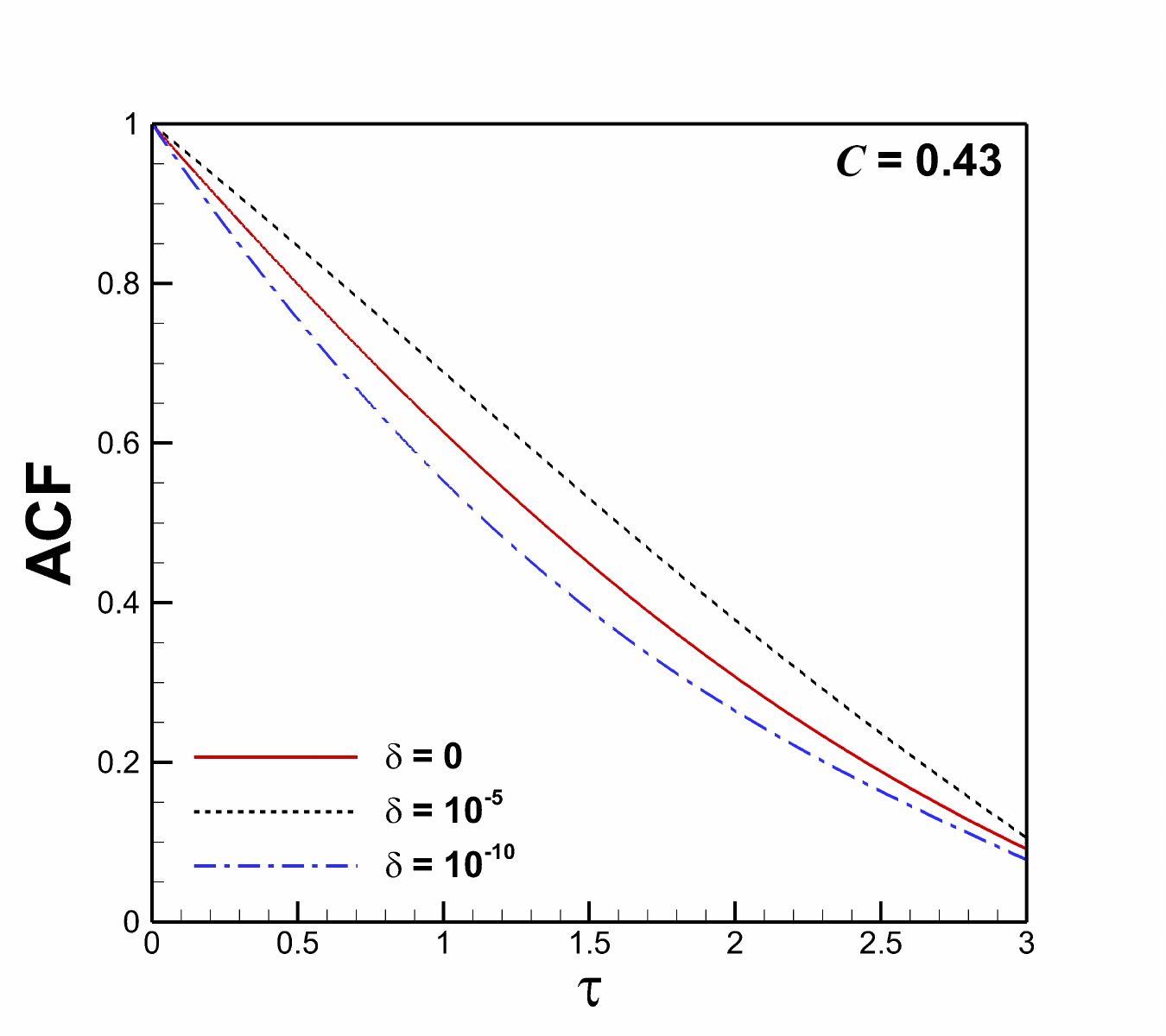}}
        \caption{Influence of tiny disturbances on the autocorrelation function (ACF) of $z(t)$ of normal-chaotic or ultra-chaotic motion of a fluid particle in ABC flow. The ACFs are based on CNS results in $t\in[0,10000]$ of a normal-chaotic or an ultra-chaotic fluid particle in ABC flow (\ref{ABC}) with (\ref{ABC-ini}) for $A=1$, $B=0.7$ and either $C=0.42$ or $C=0.43$ from the starting point ${\bf r}_{0}'=(0,0,0)+\hspace{0.2mm}(0,0,1) \times \delta$ when $\delta=0$ (red), $\delta=10^{-5}$ (black), and $\delta=10^{-10}$ (blue), respectively. (a) Variation in ACF with $\tau$ of normal-chaotic particle when $C=0.42$, and (b) variation in ACF with $\tau$ of ultra-chaotic particle when $C=0.43$, where $\tau$ denotes the lag.}    \label{ACF}
    \end{center}
\end{figure}

Let us further investigate some other statistics such as the variance $\sigma^2$, the kurtosis $\gamma_2$, and the autocorrelation function (ACF) to demonstrate the higher disorder of the ultra-chaos than the normal-chaos mentioned above, as shown in table~\ref{moments-C042} for the normal-chaos and table~\ref{moments-C043} for the ultra-chaos. Obviously, the statistics of the normal-chaos for $C=0.42$ (see table~\ref{moments-C042}) are stable to small disturbances. Conversely, the statistics of ultra-chaos for $C=0.43$ (see table~\ref{moments-C043}) are sensitive to tiny disturbances and thus unstable. In addition, the autocorrelation function (ACF) of the ultra-chaos is also sensitive to small disturbances, compared with that of the normal-chaos, as shown in figure~\ref{ACF}.

\begin{figure}
    \begin{center}
            \subfigure[]{\includegraphics[width=2.55in]{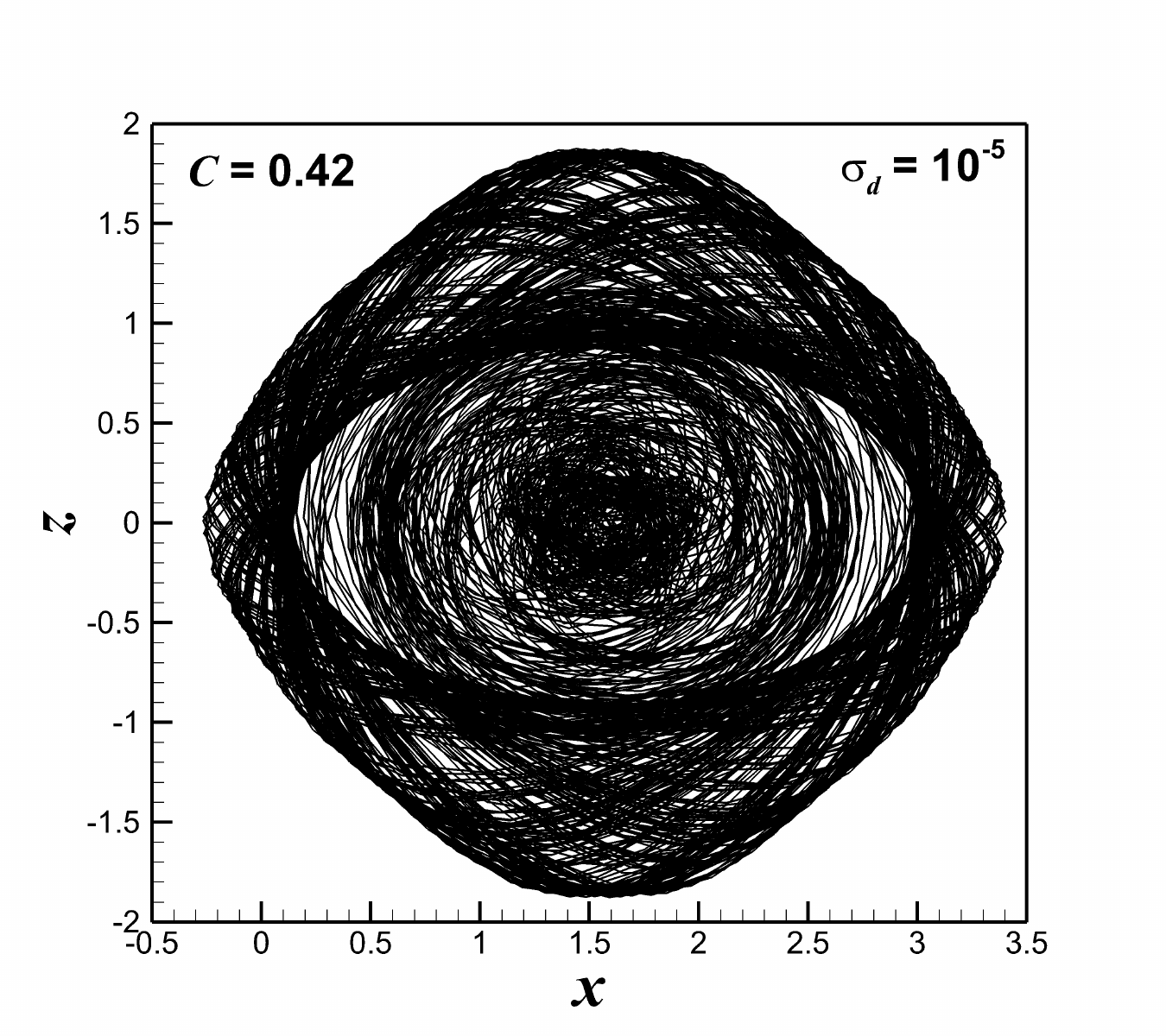}}
            \subfigure[]{\includegraphics[width=2.55in]{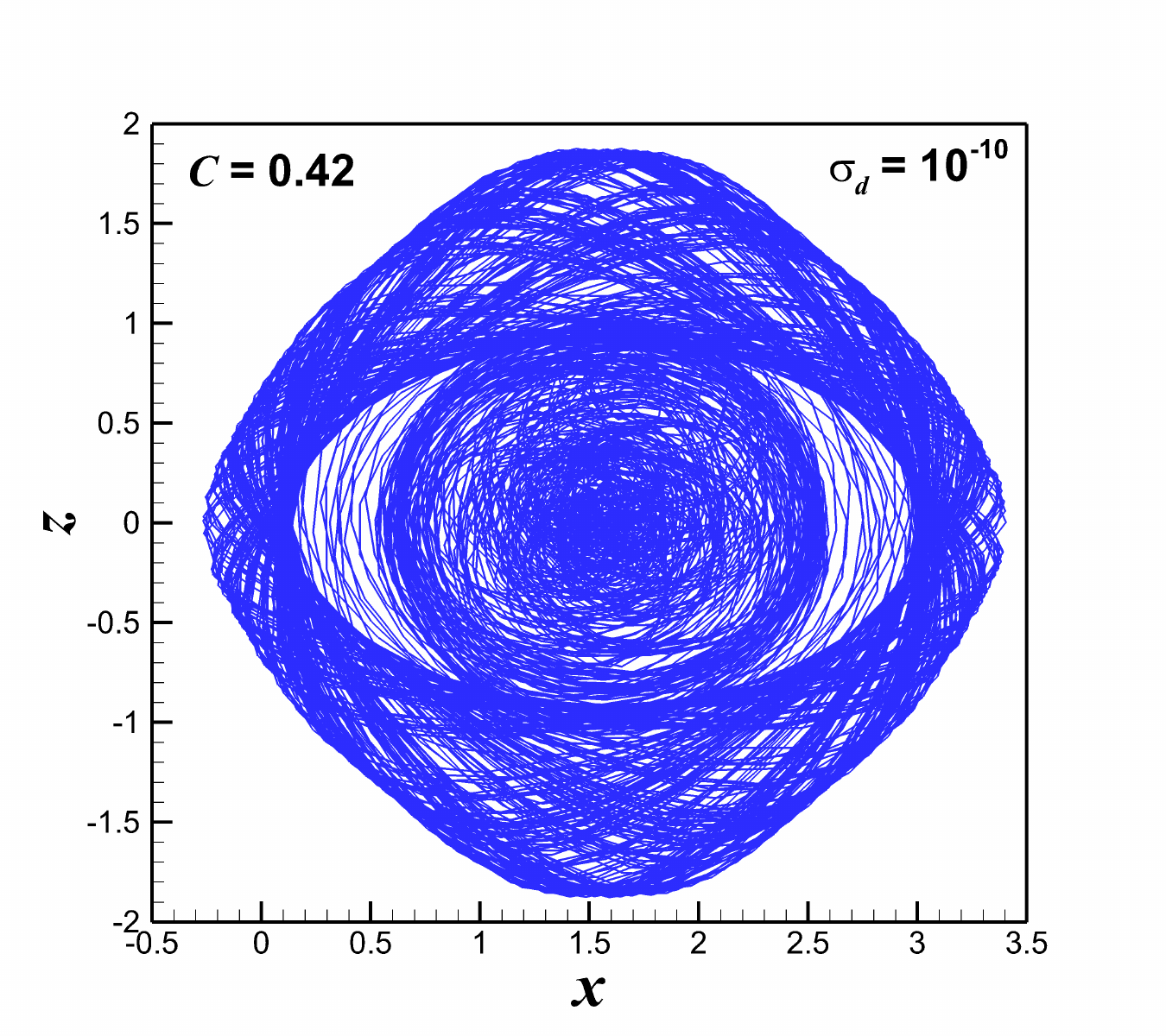}}\\
            \subfigure[]{\includegraphics[width=3.825in]{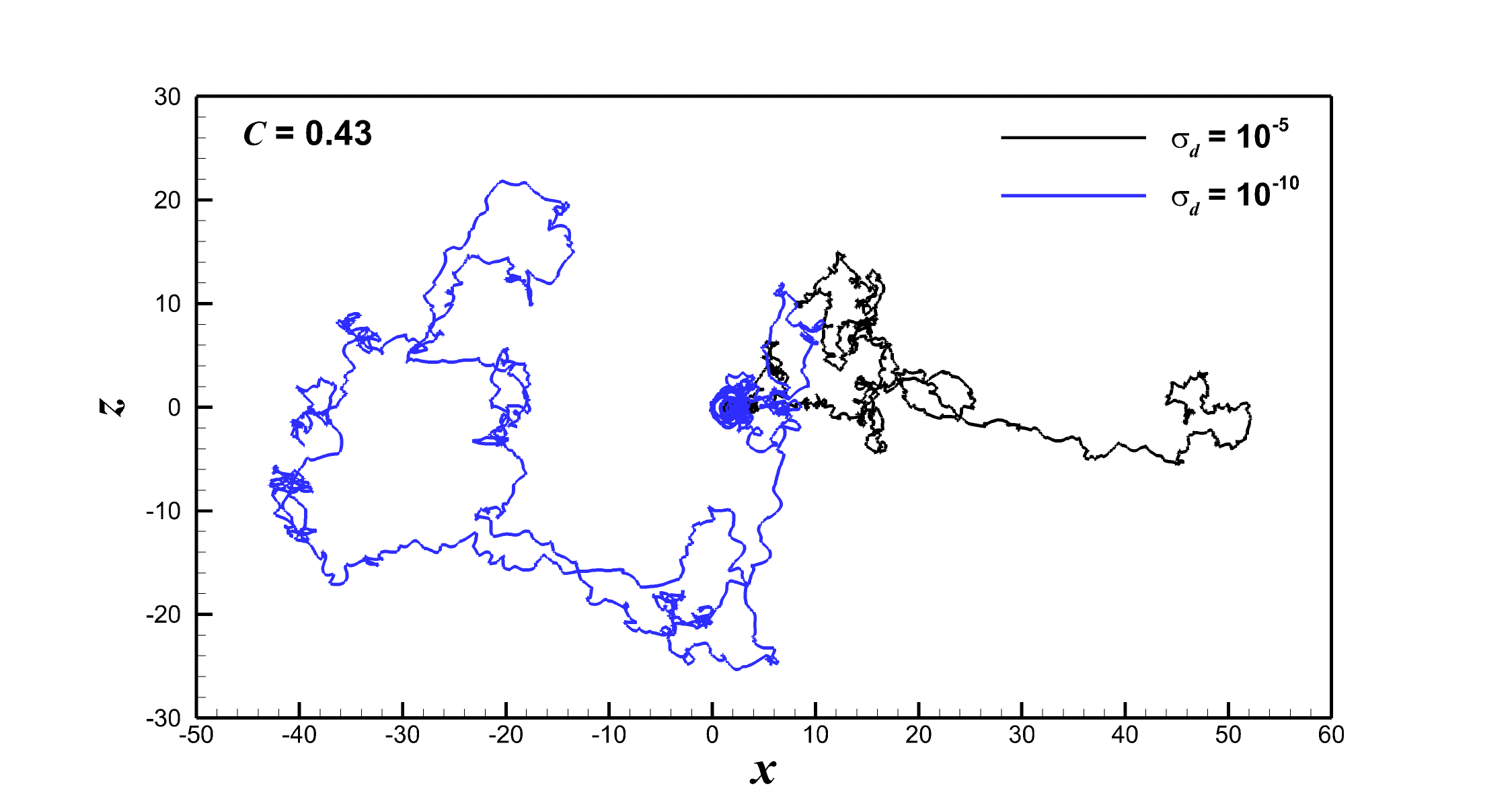}}
        \caption{Influence of tiny disturbances on the $x-z$ phase plot of ensemble-averaged trajectory of a normal-chaotic ($C=0.42$)  or an ultra-chaotic ($C=0.43$) fluid particle in ABC flow (\ref{ABC}) for $A=1$ and $B=0.7$. They are based on CNS results in $t\in[0,10000]$ from the starting point ${\bf r}_{0}=(0,0,0)+(0,0,1)\times \delta_i $,  $1\leq i\leq1000$, with $\sigma_{d}=\sqrt{\langle\delta^{2}_{i}\rangle}=10^{-5}$ (black) and $\sigma_{d} = 10^{-10}$ (blue), respectively. (a) The $x$-$z$ phase-plot of the normal-chaotic fluid particle when $C=0.42$ with $\sigma_{d}=10^{-5}$; (b) The $x$-$z$ phase-plot of the normal-chaotic fluid particle with $\sigma_{d}=10^{-10}$; (c) The $x$-$z$ phase-plot of the ultra-chaotic fluid particle when $C=0.43$ with either $\sigma_{d}=10^{-5}$ or $10^{-10}$.  }    \label{xz_ea}
    \end{center}
\end{figure}

\begin{figure}
    \begin{center}
            \subfigure[]{\includegraphics[width=2.55in]{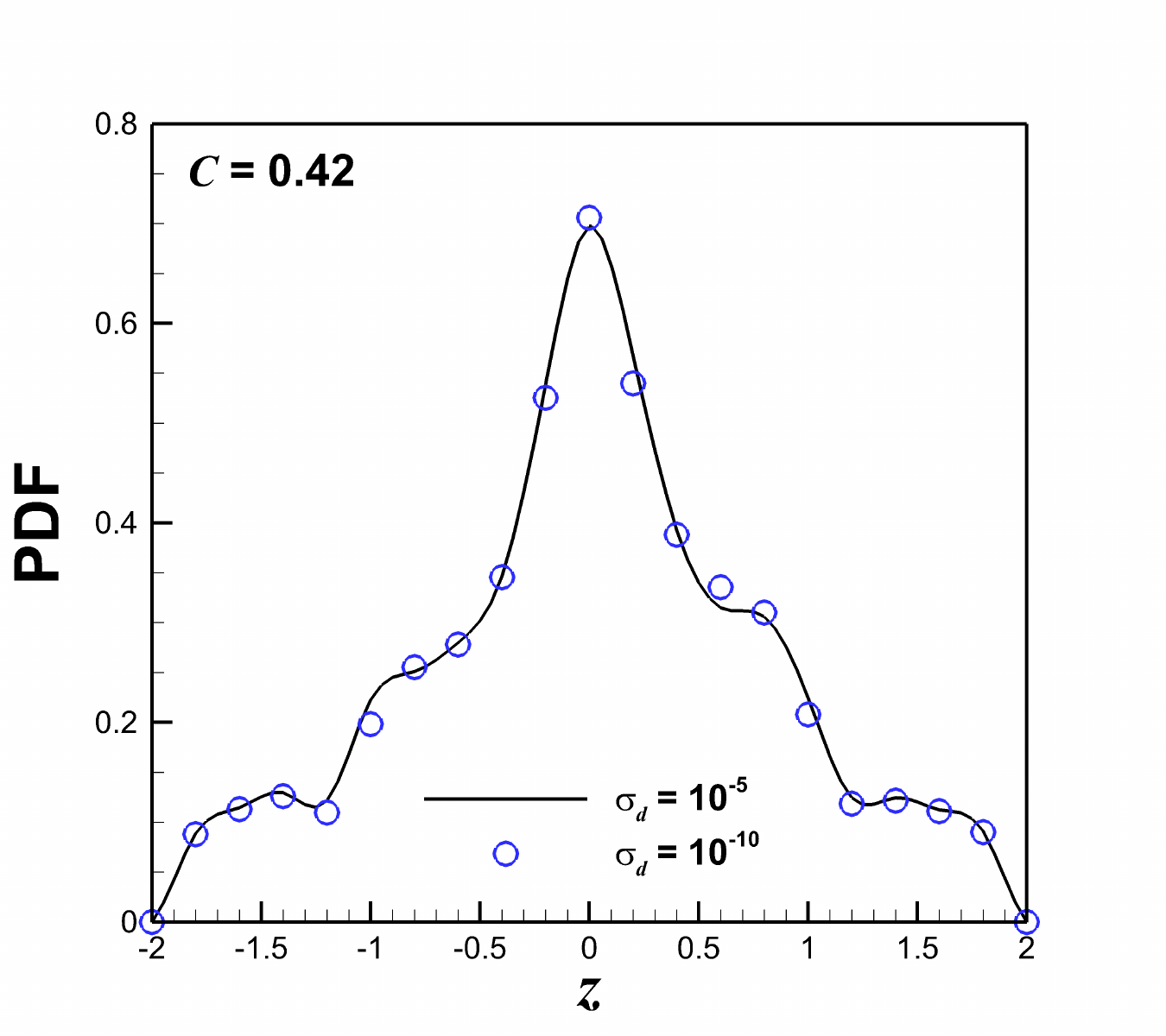}}
            \subfigure[]{\includegraphics[width=2.55in]{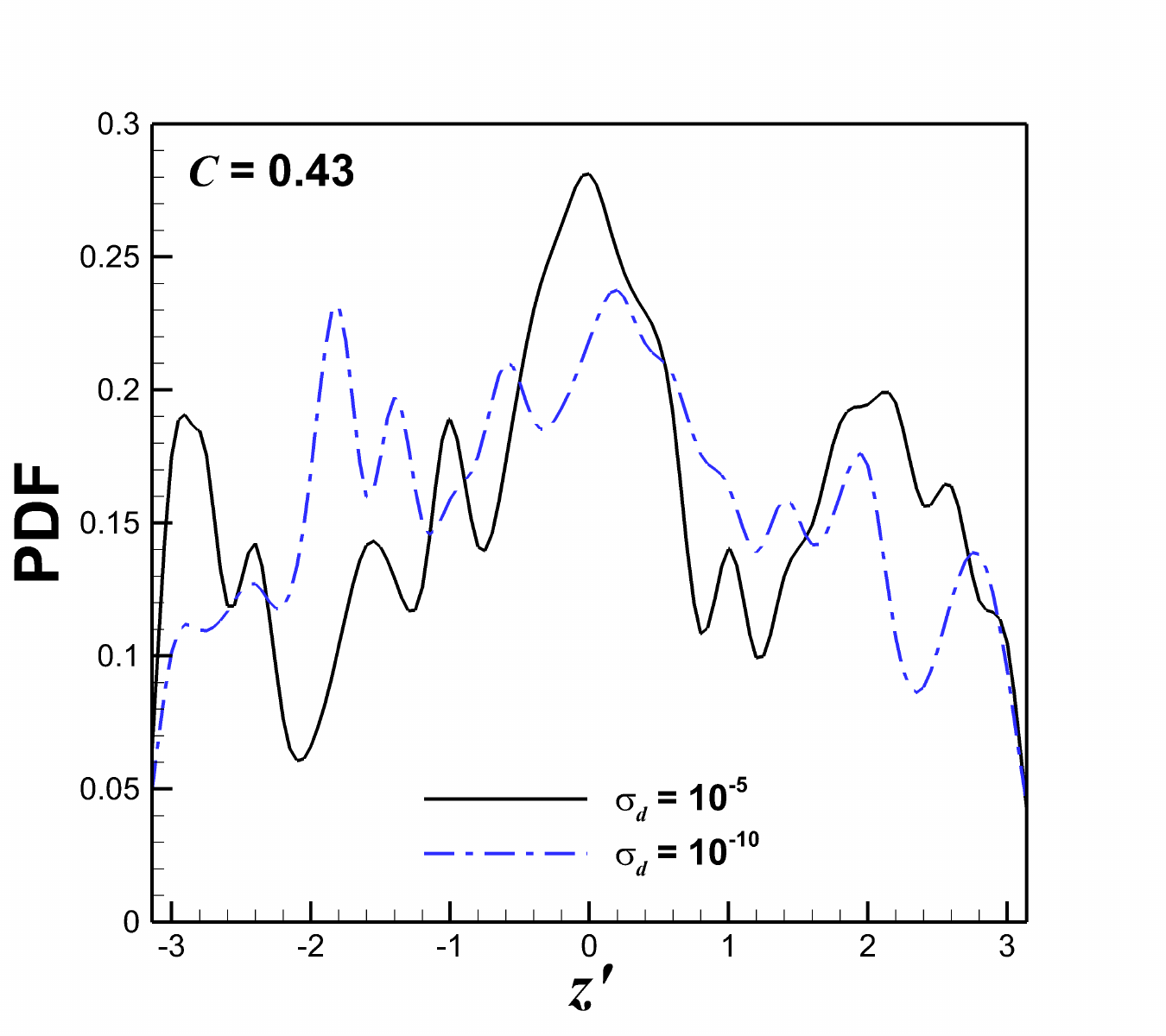}}
        \caption{Influence of tiny disturbances on the PDF of ensemble-averaged trajectory of a normal-chaotic ($C=0.42$) or an ultra-chaotic ($C=0.43$) fluid particle in ABC flow (\ref{ABC}) with (\ref{ABC-ini}) for $A=1$ and $B=0.7$. The PDFs of the ensemble-averaged trajectories are based on the CNS results in $t\in[0,10000]$ from the starting point ${\bf r}_{0}=(0,0,0)+(0,0,1) \times \delta_i $, $1\leq i\leq 1000$, with $\sigma_{d}=\sqrt{\langle\delta^{2}_{i}\rangle}=10^{-5}$ (black) and $\sigma_{d} = 10^{-10}$ (blue), respectively. (a) The PDFs of $z(t)$ of the normal-chaotic fluid particle when $C=0.42$; (b) The PDFs of the normalized results $z'(t)$ of the ultra-chaotic fluid particle when $C=0.43$.}    \label{PDF_ea}
    \end{center}
\end{figure}

Furthermore, let us consider the ensemble average of chaotic trajectories of a fluid particle starting from the point ${\bf r}_{0}' = {\bf r}_{0}+ (0,0,1) \times \delta_{i}$ with 1000 different tiny disturbances $\delta_{i}$ ($i=1,2,3, ...,1000$), which are given by the Gaussian random number generator with a standard deviation $\sigma_{d}=\sqrt{\langle\delta^{2}_{i}\rangle}$ and a zero mean, i.e. $\mu_d =\langle\delta_{i}\rangle= 0$, where $\langle\;\rangle$ denotes the average operator. For $A=1$, $B=0.7$ and $C=0.42$ and ${\bf r}_{0} = (0,0,0)$, corresponding to the normal-chaotic motions of a fluid particle, the ensemble averages of the phase plots $x-z$, which are given respectively either by $\sigma_{d}=10^{-5}$ or $\sigma_{d}=10^{-10}$, are almost the same, as shown in figure~\ref{xz_ea}(a) and (b).
On the contrary, for $A=1$, $B=0.7$ and $C=0.43$, the ensemble averages of the phase plots $x-z$ (of the ultra-chaotic motions of a fluid particle), which are given by either $\sigma_{d}=10^{-5}$ or $\sigma_{d}=10^{-10}$, are totally different, as shown in figure~\ref{xz_ea}(c).
Furthermore, the PDF of ensemble-averaged trajectory of the ultra-chaotic fluid particle starting from ${\bf r}'_{0}$ is also very sensitive to the standard deviation $\sigma_{d}$ of the starting position, which is completely different from that given by the normal-chaotic fluid particle, as illustrated in figure~\ref{PDF_ea}. These results indicate that, unlike a normal-chaos, even ensemble-averaged quantities and their corresponding PDFs for an ultra-chaos in ABC flow are unstable, i.e. rather sensitive to tiny disturbances.  Indeed, the ultra-chaotic motion is  at a higher level of disorder than that of a normal-chaos in ABC flow.

It should be emphasized that the main characteristic of ultra-chaos is that some statistics such as PDF are extremely sensitive to tiny disturbances \citep{Liao2022AAMM}. Thus, in this paper we analyze the influence of tiny disturbances in starting position on the chaotic motions of fluid particles in ABC flow. According to the results mentioned above, other statistical properties (such as variance, kurtosis, ACF, ensemble-averaged trajectory, and ensemble-averaged trajectory's PDF) of ultra-chaos in the ABC flow are also very sensitive to tiny disturbances in the starting position. On the contrary, these statistics, given by CNS results of a normal-chaotic fluid particle in the ABC flow, are not sensitive to tiny disturbances. This indicates that the statistics of a normal-chaos are stable.

\begin{figure}
    \begin{center}
            \includegraphics[width=2.55in]{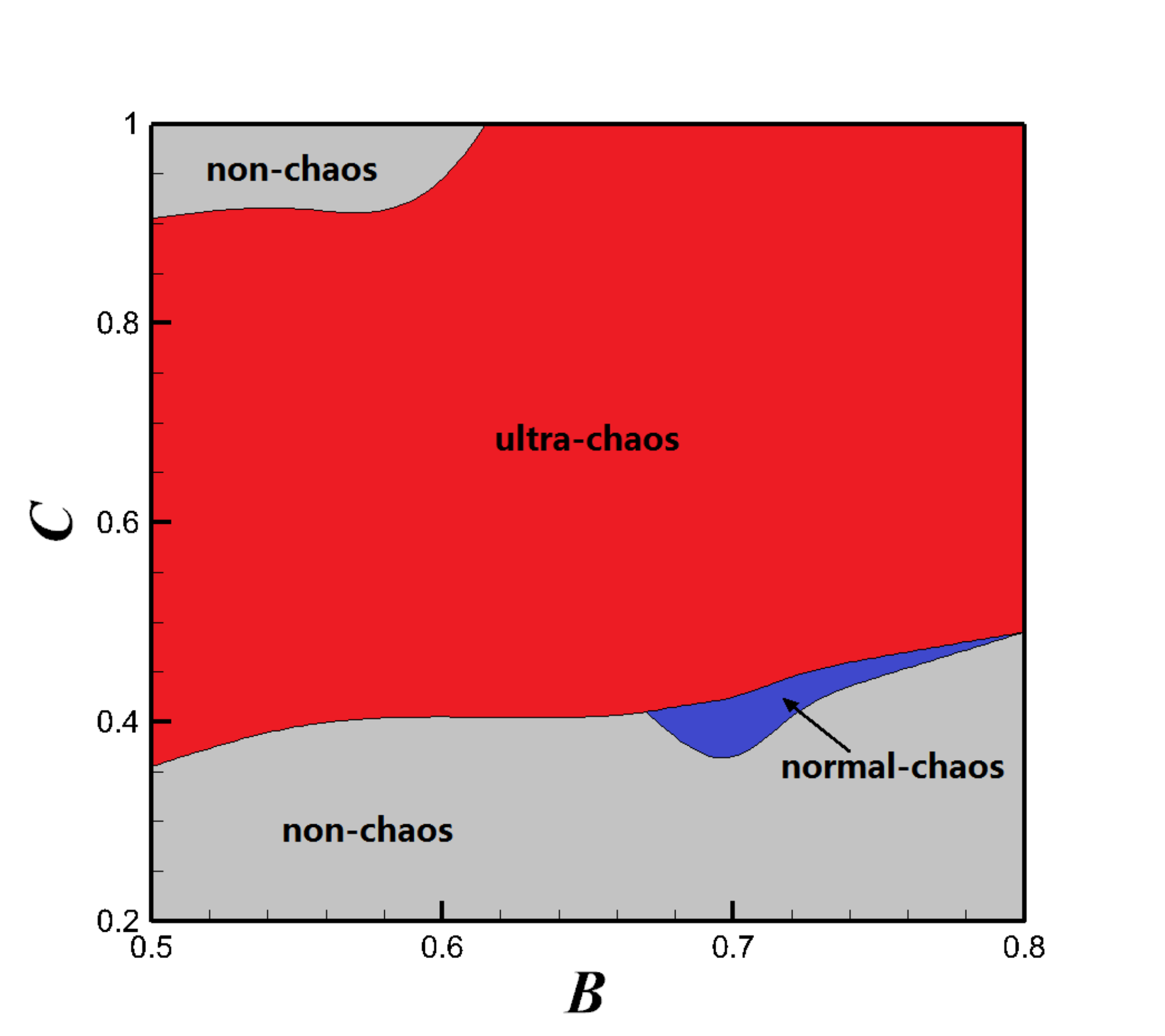}
        \caption{Classification of trajectories of fluid particles in ABC flow (\ref{ABC}) starting from ${\bf r}_{0} = (0,0,0)$ for different values of $B$ and $C$ when $A=1$. Gray domain: non-chaos; Blue domain: normal-chaos; Red domain: ultra-chaos. }    \label{ultra}
    \end{center}\vspace{-0.6cm}
\end{figure}

For $A=1$ and varying the values of $B$ and $C$, we find that non-chaos, normal-chaos, and ultra-chaos all exist in the fluid particle trajectory starting from ${\bf r}_{0} = (0,0,0)$, as shown in figure~\ref{ultra}. For a non-chaotic motion, the fluid particle trajectory is {\em stable} to tiny disturbances in the starting position. For a normal-chaotic motion, although the trajectory is rather sensitive  to tiny disturbances in the starting position, i.e. {\em unstable}, the phase plot and the statistical properties are {\em stable} to the tiny disturbances. However, for an ultra-chaotic motion, even the statistical properties are {\em unstable}, say, sensitive to tiny disturbances in the starting position. 
As we can see in figure~\ref{ultra}, for $A=1$ and $B=0.7$, a small change between $C=0.42$ and $C=0.43$ triggers the transition from normal-chaotic motion to ultra-chaotic motion, which is the reason why we present the cases in figure~\ref{C042} and figure~\ref{C043}.
Note that for the normal-chaotic motion, the fluid particle starting from ${\bf r}_{0}=(0,0,0)$ always moves in a restricted spatial domain (such that its position is in a restricted domain of the phase plot $x-z$ as shown in figure~\ref{C042}). However, for an ultra-chaotic motion, the fluid particle starting from ${\bf r}_{0}=(0,0,0)$ progressively departs from its starting point, further illustrating that ultra-chaotic motion in ABC flow has higher disorder than normal-chaotic motion, although the velocity field of the ABC flow as a whole is inherently periodic and steady-state.

\begin{figure}
    \begin{center}
            \subfigure[]{\includegraphics[width=1.7in]{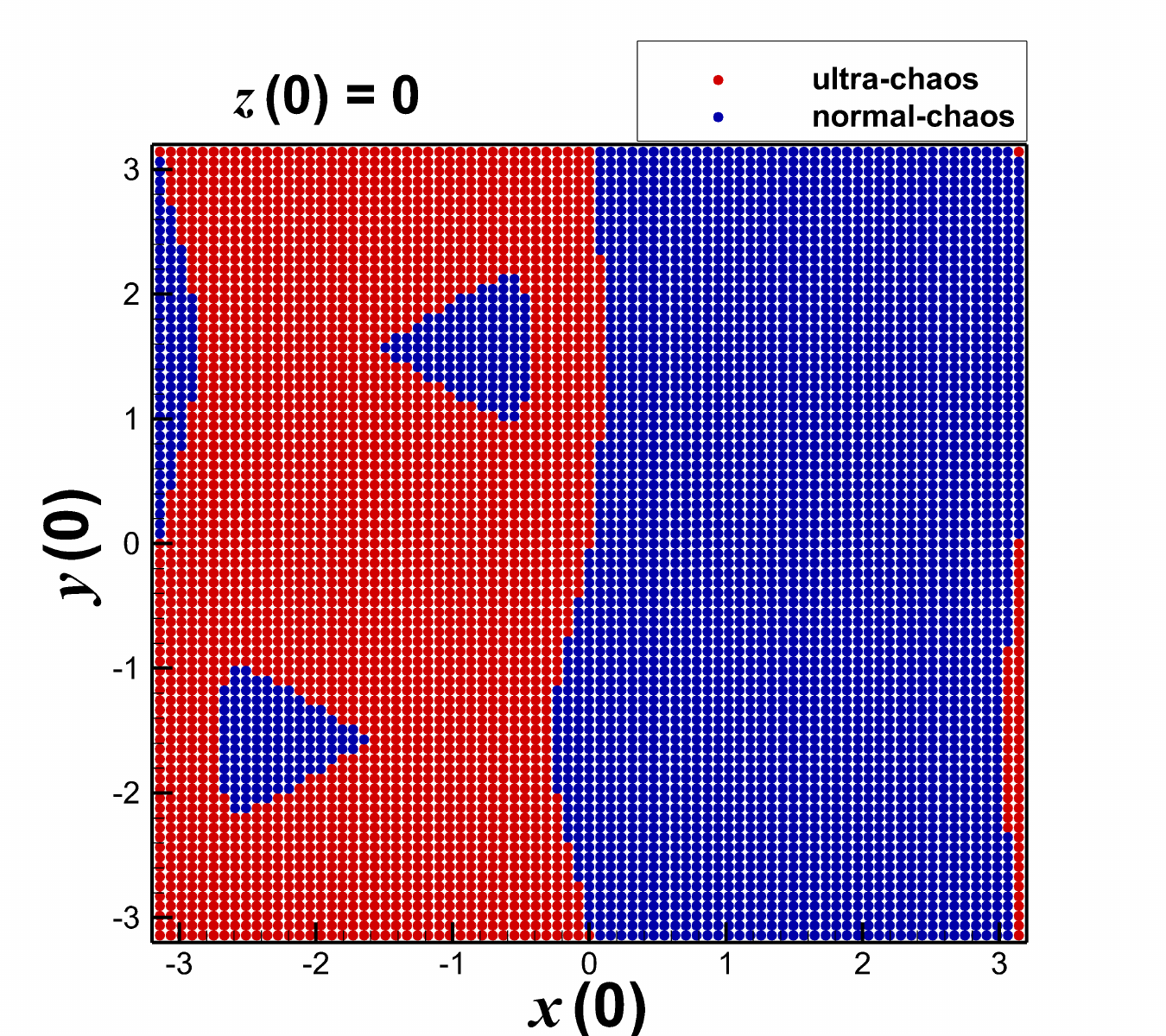}}
            \subfigure[]{\includegraphics[width=1.7in]{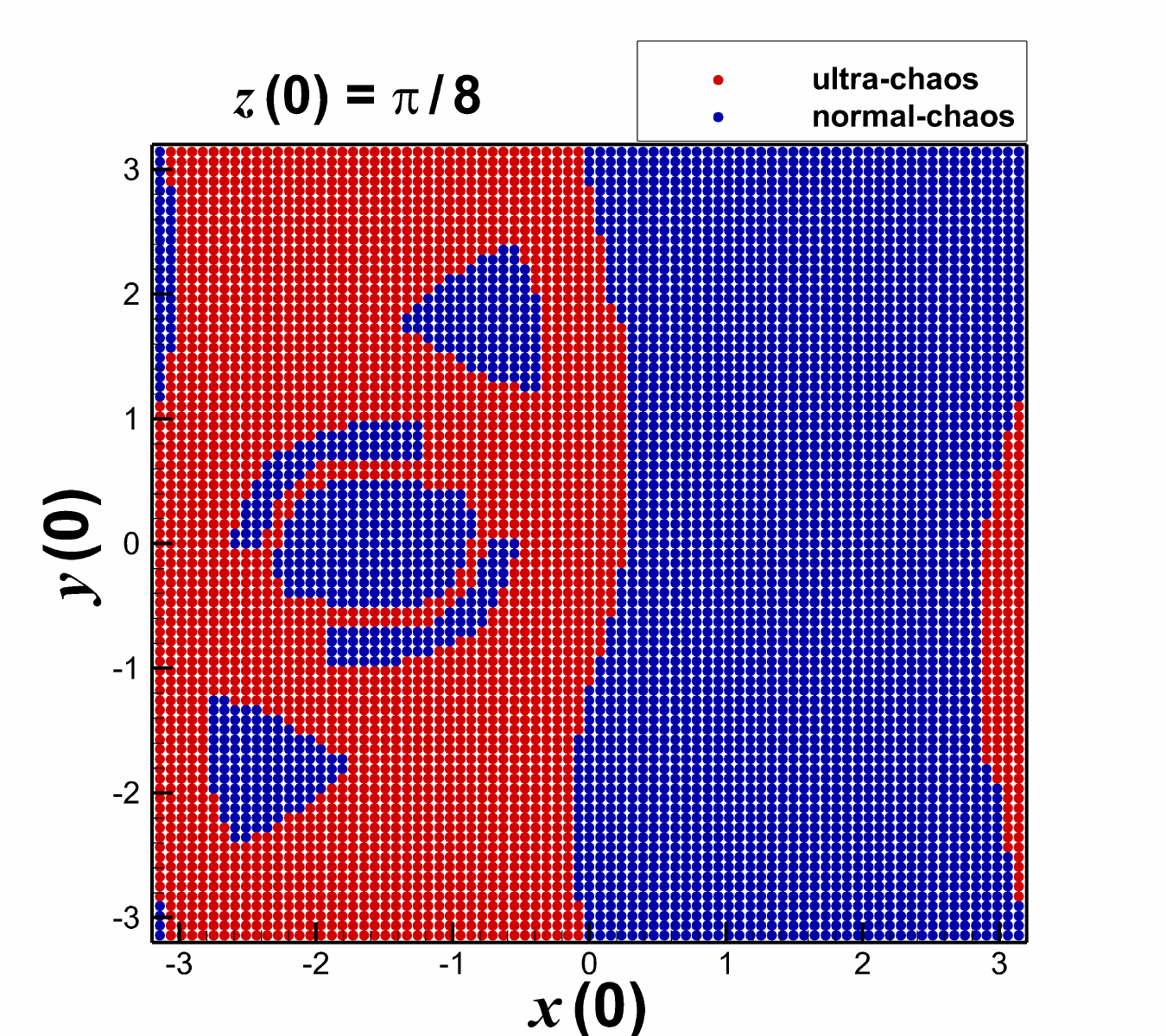}}
            \subfigure[]{\includegraphics[width=1.7in]{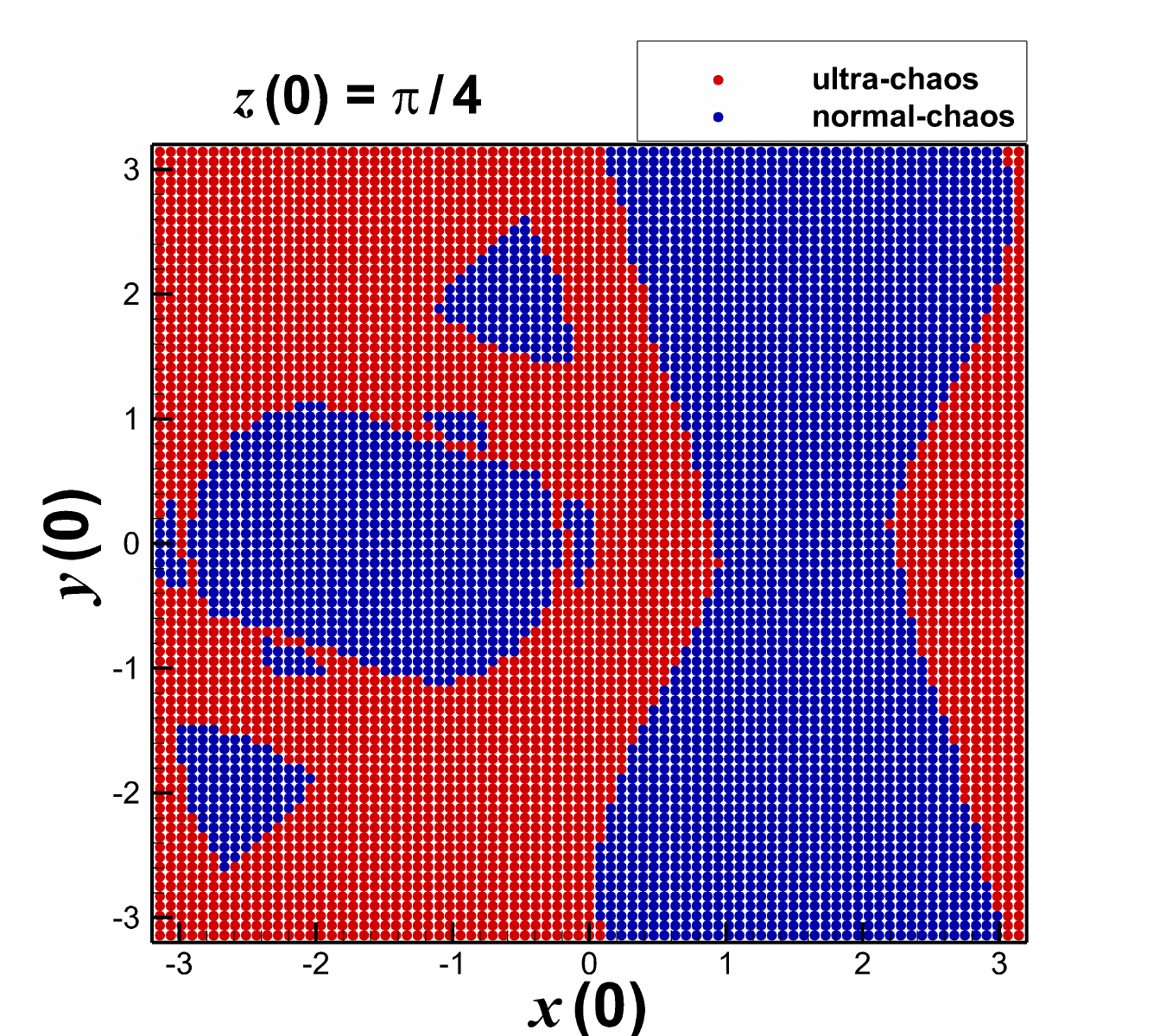}}
            \subfigure[]{\includegraphics[width=1.7in]{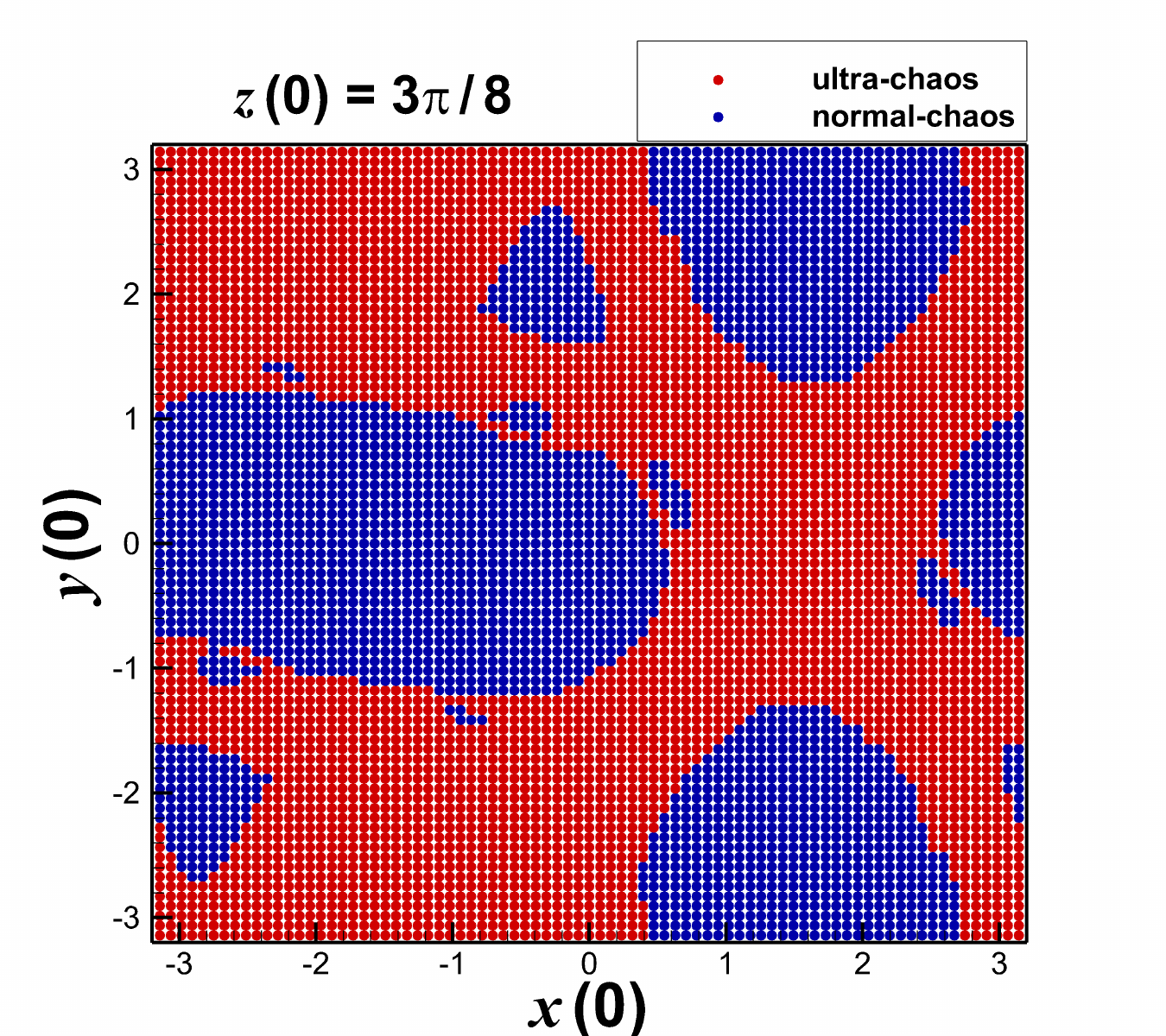}}
            \subfigure[]{\includegraphics[width=1.7in]{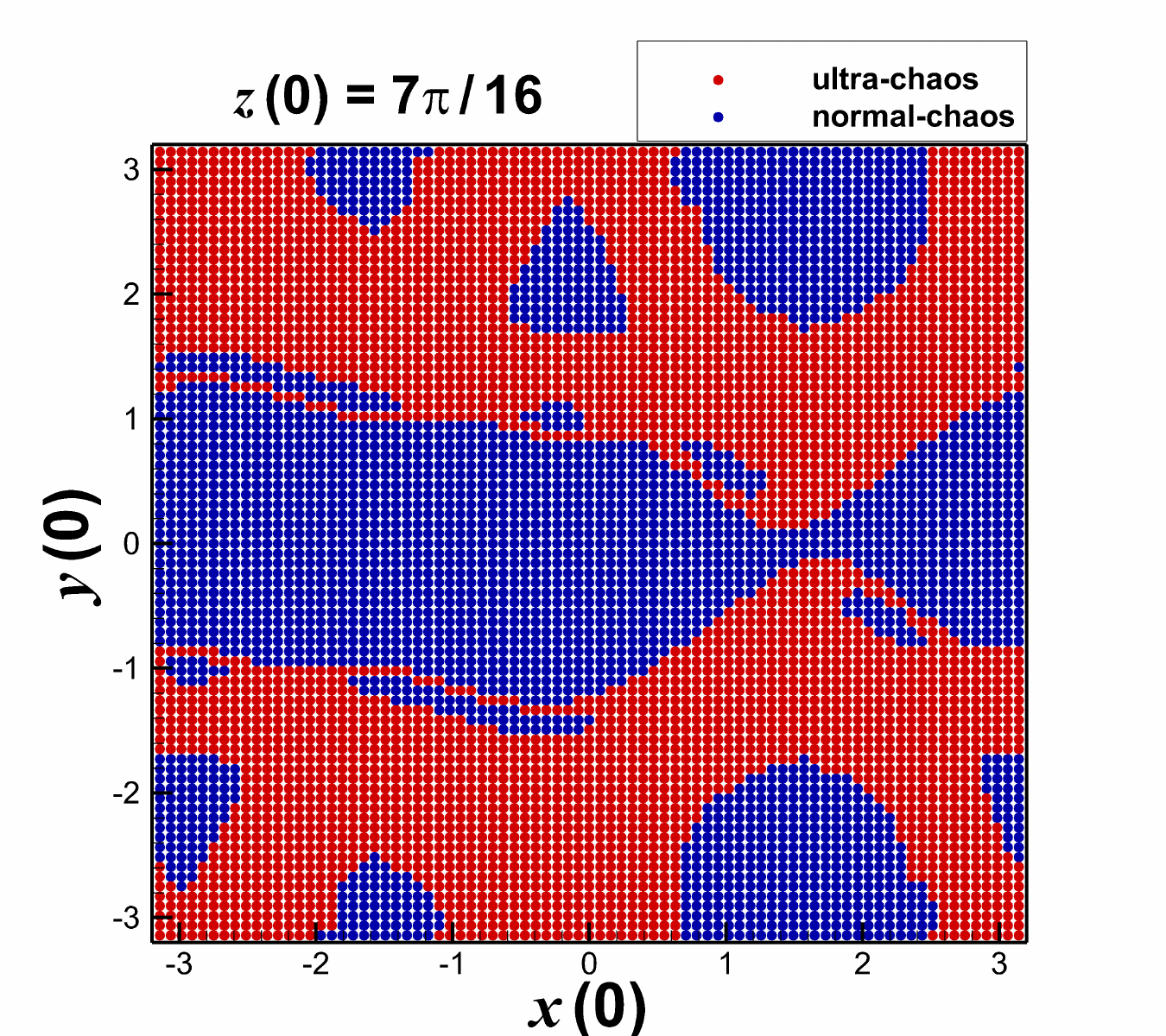}}
            \subfigure[]{\includegraphics[width=1.7in]{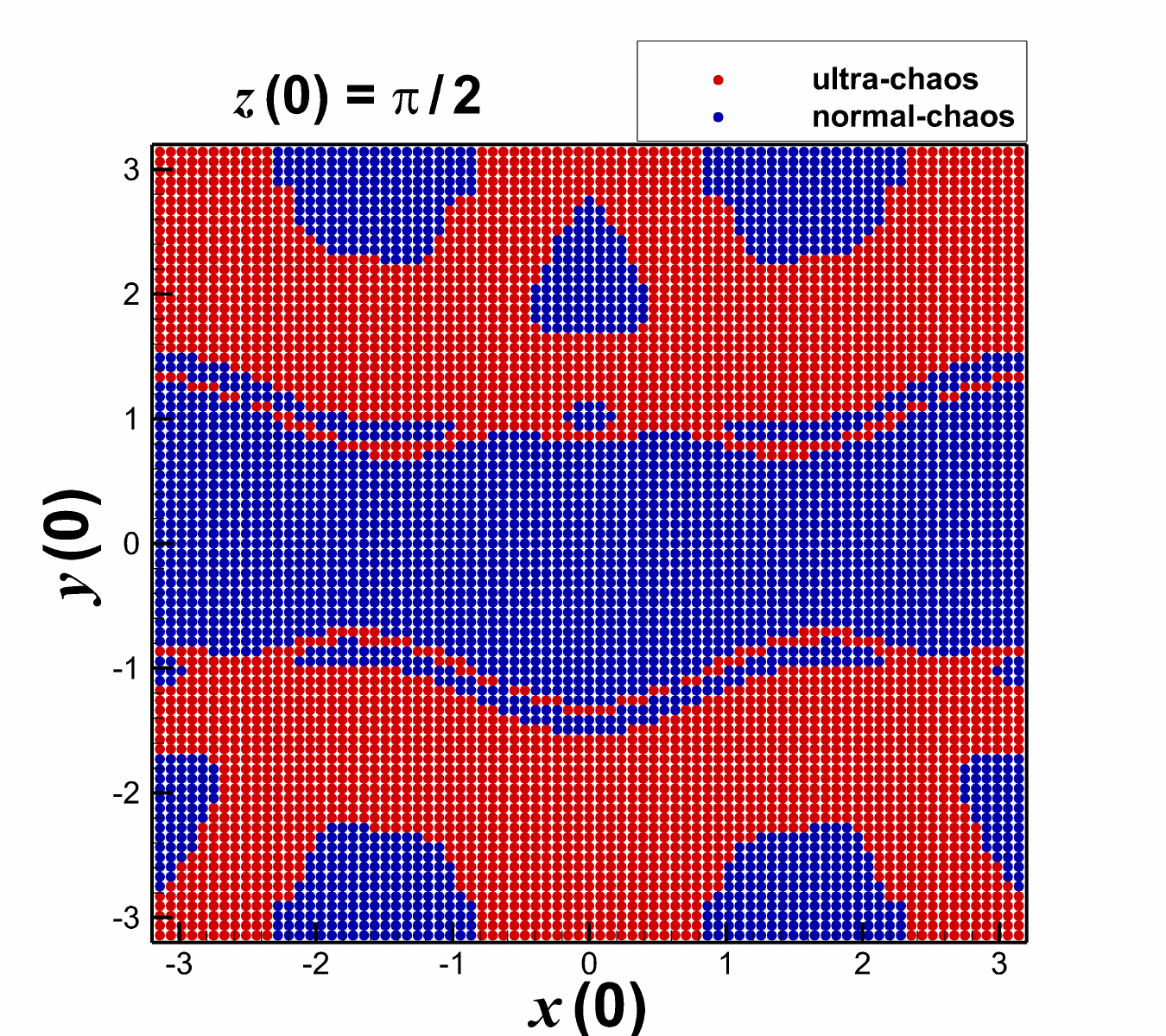}}
        \caption{Chaotic states of trajectories of the fluid particles starting from different points ${\bf r}_{0}=(x(0),y(0),z(0))$ in ABC flow (\ref{ABC}) for $A=1$, $B=0.7$ and $C=0.43$. (a) $z(0)=0$; (b) $z(0)=\pi/8$; (c) $z(0)=\pi/4$; (d) $z(0)=3\pi/8$; (e) $z(0)=7\pi/16$; (f) $z(0)=\pi/2$. Blue points: normal-chaos; Red points: ultra-chaos.}    \label{r0_1}
    \end{center}\vspace{-0.6cm}
\end{figure}

\begin{table}
\tabcolsep 0pt
\vspace*{-2pt}
\begin{center}
\begin{tabular}{ccc}
~~$~$ & \hspace{1.0cm}  Normal-chaos \hspace{1.0cm}  & \hspace{1.0cm}  Ultra-chaos\hspace{1.0cm}   \\
\\
~~Maximum value of $\lambda_{max}$ & $1.3\times 10^{-2}$ & $8.7\times 10^{-2}$~~ \\
~~Minimum value  of $\lambda_{max}$ & $8.5\times 10^{-5}$ & $4.3\times 10^{-2}$~~ \\
~~Mean of $\lambda_{max}$ & $9.7\times 10^{-4}$ & $6.9\times 10^{-2}$~~ \\
~~Standard deviation of $\lambda_{max}$ & $7.5\times 10^{-4}$ & $1.0\times 10^{-2}$~~ \\
\end{tabular}
\end{center}
\caption{Statistical values of maximum Lyapunov exponents $\lambda_{max}$ of the normal-chaotic  and ultra-chaotic trajectories of fluid particles in the ABC flow, gained by solving the chaotic dynamic system ABC flow (\ref{ABC}) in $t\in[0,10000]$ for $A=1$, $B=0.7$ and $C=0.43$ by means of CNS, using various starting points ${\bf r}_{0}=(x(0),y(0),z(0))$ of the fluid particles, where $-\,\pi \leq x(0),y(0),z(0)\leq +\,\pi$.}    \label{Lye}
\end{table}

On the other hand, keeping $A=1$, $B=0.7$ and $C=0.43$ and using various positions of the starting point ${\bf r}_{0}=(x(0),y(0),z(0))$ in ABC flow, where $-\,\pi \leq x(0),y(0),z(0)\leq +\,\pi$, it is found in a similar way that both normal-chaos and ultra-chaos (for the motions of fluid particles starting from different ${\bf r}_{0}$) widely exist, and these two states of chaos co-exist simultaneously in the ABC flow, as shown in figure~\ref{r0_1}.
The statistical values of their maximum Lyapunov exponents $\lambda_{max}$ are given in table~\ref{Lye}. Statistically speaking, the maximum Lyapunov exponents $\lambda_{max}$ of the ultra-chaotic motions of fluid particles in ABC flow are about two orders of magnitude larger than those of the normal-chaos.

Note that, when $z(0)$ of the starting point ${\bf r}_{0}=(x(0),y(0),z(0))$  increases from $0$ to $\pi/2$, there exists a kind of structure constituted by the starting positions $(x(0),y(0))$ of fluid particles with normal-chaotic motion (corresponding to blue points) and ultra-chaotic motion (red points), which undergoes continuous deformation, as shown in figure~\ref{r0_1}. 
Although normal-chaotic motion is qualitatively different from ultra-chaotic motion so that it is not difficult for us to obtain the general structure in figure~\ref{r0_1}, we still require a criterion by which to quantitatively determine the boundary of the structure. Let $f(z')$ denote the PDF of a normalized result $z'\in[-\pi,+\pi)$ given by a chaotic motion of fluid particle, and $f^*(z')$ the PDF of  another one with a tiny disturbance to the starting position. A criterion based on the following relative error  
\begin{equation}
\frac{\int_{-\pi}^{+\pi}\mid f(z')-f^*(z')\mid dz'}{\int_{-\pi}^{+\pi}f(z')\,dz'}=\int_{-\pi}^{+\pi}\mid f(z')-f^*(z')\mid dz' \leq \gamma 
\end{equation}
is usually adopted to determine the boundary between normal-chaos and ultra-chaos, as shown in figure~\ref{r0_1}. According to our experience, $\gamma=5\%$ is often suitable to distinguish between a normal-chaotic motion and an ultra-chaotic motion.
Figure~\ref{r0_1} illustrates that the normal-chaotic and ultra-chaotic states co-exist at the same time, which is reasonable in a volume-preserving ABC flow that has different types of chaotic trajectories for the motions of fluid particles \citep{dombre1986chaotic}, which will be discussed later in detail.

Let $\alpha(x(0),y(0),z(0))=0$ or $1$ denote either a normal-chaotic motion or an ultra-chaotic motion  of a fluid particle starting from ${\bf r}_{0}=(x(0),y(0),z(0))$, respectively. Then, according to our computations, for $-\,\pi \leq x(0) \leq +\,\pi$, there exist the symmetries
\begin{equation}
\alpha(x(0),y(0),z(0))=\alpha(-\,x(0),y(0),\pi-z(0)),  
\end{equation}
where $ y(0) \in [-\,\pi, +\,\pi]$, $z(0)\in [\pi/2, \pi]$,
\begin{equation}
\alpha(x(0),y(0),z(0))=\alpha(x(0),\pi-y(0),-\,z(0)),
\end{equation}
where $y(0)\in[0, \pi]$, $z(0)\in[-\,\pi, 0]$, and
\begin{equation}
\alpha(x(0),y(0),z(0))=\alpha(x(0),-\,\pi-y(0),-\,z(0)),
\end{equation}
where $y(0)\in[-\,\pi, 0]$, $z(0)\in[-\,\pi, 0]$, respectively.  

Considering the fact that the normal-chaotic and ultra-chaotic states co-exist simultaneously as shown in figure~\ref{r0_1}, without loss of generality, we choose two starting points ${\bf r}_{0,n}=(0,-0.1,0)$ and ${\bf r}_{0,u}=(-0.1,0.1,0)$ of fluid particles in ABC flow to illustrate a normal-chaotic motion (left) and an ultra-chaotic motion (right) via a movie (see the supplementary movie) in the case of $A=1, B =0.7, C=0.43$ within $t\in[0,5000]$. As shown in the left part of the movie (corresponding to a normal-chaos), the fluid particle starting from ${\bf r}_{0,n}$ always moves in a restricted spatial domain and the corresponding trajectory resembles weak chaos.     
On the contrary,  the fluid particle starting from ${\bf r}_{0,u}$ departs the starting point far away and even its position normalized by periodic condition appears to be in disorder, as shown in the right part of the movie (corresponding to ultra-chaos).       
All of these clearly illustrate that an ultra-chaotic motion in  the ABC flow is completely different from a normal-chaotic motion: an ultra-chaos has indeed a much higher disorder than a normal-chaos.  

\begin{table}
\tabcolsep 0pt
\vspace*{-2pt}
\begin{center}
\begin{tabular}{lc}
 $C$ \hspace{1.0cm} & \hspace{1.0cm} $N_{ultra}\hspace{0.2mm}/\hspace{0.2mm}N_{all}$\hspace{1.0cm} \\
\\
~~$0.43$ & $49\%$~~ \\
~~$0.2$ & $47\%$~~ \\
~~$0.1$ & $43\%$~~ \\
~~$0.01$ & $20\%$~~ \\
~~$0.001$ & $6\%$~~ \\
~~$0.0001$ & $2\%$~~ \\
~~$0$ & $0\%$~~ \\
\end{tabular}
\end{center}
\caption{Values of the parameter $C$ versus $N_{ultra}\hspace{0.2mm}/\hspace{0.2mm}N_{all}$, where $N_{ultra}$ denotes the number of starting points corresponding to an ultra-chaotic trajectory of fluid particle in the ABC flow and $N_{all}$ denotes the total number of equidistant starting points. The results are obtained by solving the chaotic dynamical system (\ref{ABC}) in $t\in[0,10000]$ for $A=1.0$, $B=0.7$ and $0 \leq C \leq 0.43$ by means of CNS, using various starting points ${\bf r}_{0}=(x(0),y(0),z(0))$ of the fluid particles, where $-\,\pi \leq x(0),y(0),z(0)\leq +\,\pi$.}    \label{C_ultra-T}
\end{table}

\begin{figure}
    \begin{center}
            \includegraphics[width=2.55in]{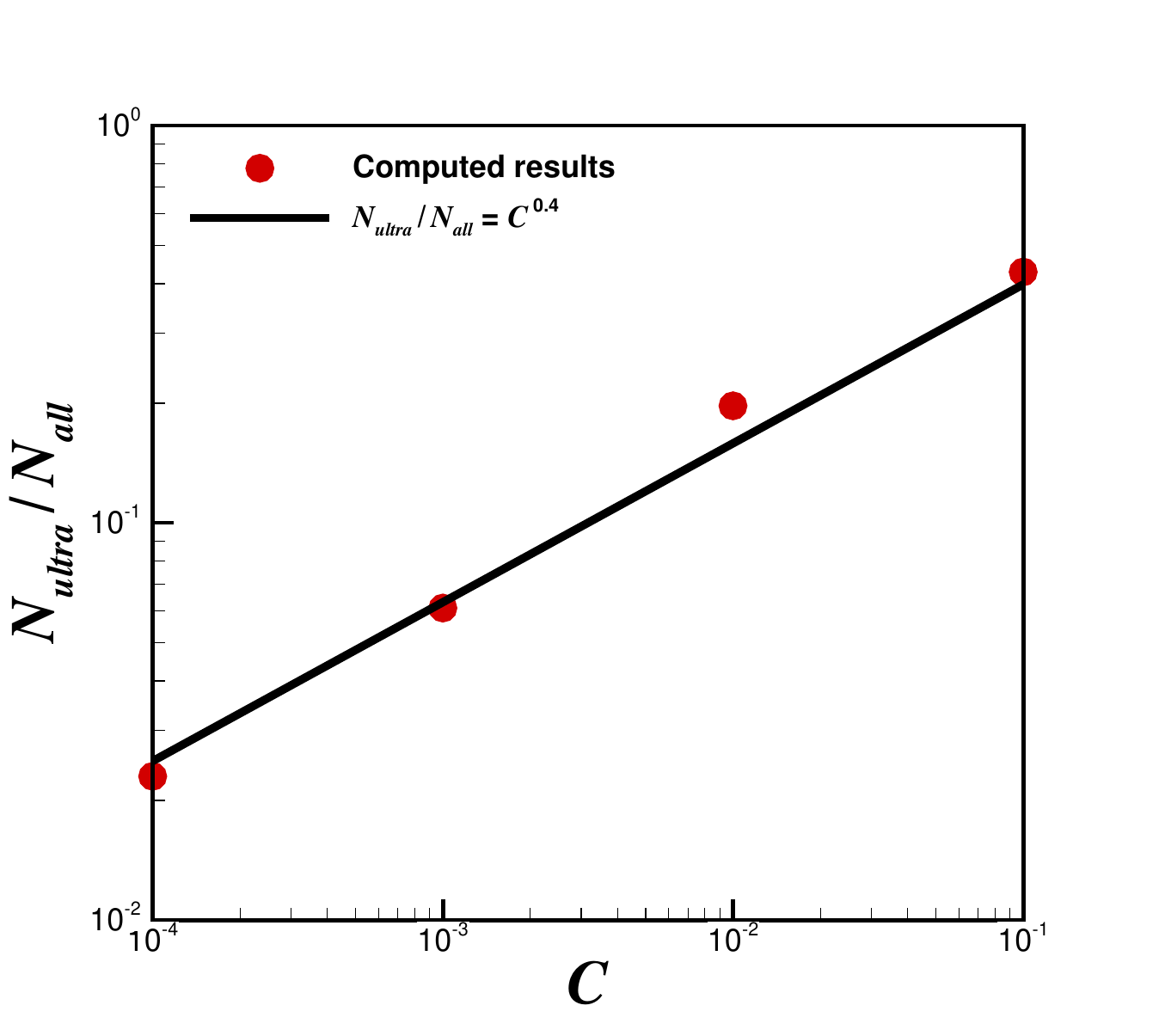}
        \caption{The ratio of the number of the starting fluid particles with ultra-chaotic motion to the total number of particles versus the value of $C$ of ABC flow for $A=1$, $B=0.7$ and $C\leq0.1$.}    \label{C_ultra-F}
    \end{center}
\end{figure}

Let $\beta$ denote the ratio of the number of the starting fluid particles with ultra-chaotic motion to the total number of particles in $-\,\pi\leq x,y,z \leq +\,\pi$. In theory
\begin{equation}
\beta = \frac{1}{(2\pi)^{3}}\int_{-\pi}^{+\pi} \int_{-\pi}^{+\pi}\int_{-\pi}^{+\pi} \alpha(x,y,z)\, dx\hspace{0.2mm}dy\hspace{0.2mm}dz,
\end{equation}
with either $\alpha=0$ for a normal-chaos or $\alpha=1$ for an ultra-chaos, respectively. In practice, we use the Monte-Carlo method to estimate the ratio
\begin{equation}
\beta \approx \frac{N_{ultra}}{N_{all}},
\end{equation}
where $N_{all}$ denotes the total number of the randomly selected starting fluid particles \[ {\bf r}_{0}\in\Omega=\Big\{(x,y,z): -\,\pi \leq x,y,z \leq +\,\pi\Big\} \]  and $N_{ultra}$ is the number of starting fluid particles with ultra-chaotic motion.  Obviously, the larger $N_{all}$, the more accurate the result of $\beta$ given by the Monte-Carlo method.
For $A=1$, $B=0.7$, $0 \leq C \leq 0.43$ and $N_{all}=8000$, it is found that the ratio $\beta$ is dependent upon the value of $C$, as shown in table~\ref{C_ultra-T}. Notably, when $C\leq0.1$, it is found that there exists a power-law relationship between $\beta$ and $C$, say, 
\begin{equation}
\beta \approx C^{\hspace{0.2mm}0.4},    \label{relation_C_ultra}
\end{equation}
as illustrated in figure~\ref{C_ultra-F}. Thus, when the parameter $C$ decreases, the value of $N_{ultra}$, i.e. the number of the starting fluid particles with ultra-chaotic motion, decreases until $N_{ultra} = 0$ when $C = 0$. This is reasonable since the ABC flow for $C=0$ is stable and thus chaotic motion of fluid particles does not exist at all in $C=0$.  

\subsection{Difference between ultra-chaos and sensitivity of statistics to parameters}

It is well known that a chaotic trajectory is unstable, i.e. sensitive to small disturbances. For normal-chaos, a trajectory is unstable but its statistics are stable to small disturbances. However, for ultra-chaos, even its statistics are unstable, i.e. sensitive to very small disturbances.  The stability of different types of dynamic system is listed in table~\ref{disorder}.  Obviously, as illustrated by many examples \citep{Liao2022AAMM, Yang2023CFS}, ultra-chaos involves higher disorder than normal-chaos. 

It should be emphasized that, unlike sensitivity to parameters,  the concept of ultra-chaos focuses on the {\em stability} of statistics of a dynamic system, while all physical parameters are {\em fixed},  to small disturbances that can be  very {\em tiny}.         
Certain dynamic systems exhibit high sensitivity in statistics to physical parameters \citep{broer2002bifurcations,  ashwin2012tipping, lucarini2019transitions, sliwiak2021computational},  which however is essentially different from ultra-chaos, i.e. instability of statistics to small disturbances. For example,  \cite{broer2002bifurcations} investigated  bifurcations and strange attractors in the Lorenz-84 climate model with seasonal forcing: 
\begin{eqnarray}
\dot{x} &=& -a x -y^{2}-z^{2} +  a F \left(1+\epsilon \cos \omega t \right), \\
\dot{y} &=& -y + x y - b x z + G \left( 1+\epsilon \cos \omega t \right), \\
\dot{z} &=& -z + b xy + xz,
\end{eqnarray}  
where $\omega = 2\pi/T$ and $a, b, T, F, G$, $\epsilon$ are physical parameters.  Without loss of generality,  \cite{broer2002bifurcations}  considered the cases of $a=1/4, b = 4, T=73$ with varying $F\in[0,12], G\in[0,9]$ and $\epsilon\in[0,0.5]$, and found that there exist Hopf bifurcations and some high sensitivity  of  statistics  to  parameters $F, G$ and $\epsilon$.  However, we found that {\em all} statistic results given by the above-mentioned Lorenz-84 climate model with seasonal forcing are {\em stable} to small disturbances, in that they are either non-chaotic or normal-chaotic.  In other words,  even when high sensitivity of statistics to physical parameters exists,  the corresponding dynamic system is stable and thus is {\em not} an ultra-chaos!   In fact, like the famous three-dimensional Lorenz equation (with one positive Lyaponov exponent) and the four-dimensional R\"{o}ssler system with two positive Lyaponov exponents) \citep{Liao2022AAMM}, the above-mentioned Lorenz-84 climate model  has  normal-chaotic trajectories at most.   This is a good example to illustrate the essential difference between ultra-chaos and high sensitivity of statistics to physical parameters: they are quite different things!    

For an ultra-chaotic system, its statistics are unstable to {\em any} types of disturbances.  For example, in the case of $A=1, B=0.7$ and $C=0.43$, the trajectory starting from (0,0,0) is still ultra-chaotic even if there is no disturbance to the starting point but a small  environmental  disturbance, governed by 
\begin{equation}
\left\{
\begin{array}{l}
\dot{x}(t)=A\sin[z(t)]+C\cos[y(t)] ,       \\
\dot{y}(t)=B\sin[x(t)]+A\cos[z(t)] ,        \\
\dot{z}(t)=C\sin[y(t)]+B\cos[x(t)] + \varepsilon(t),
\end{array}
\right.  \label{ABC2}
\end{equation}
 with the initial condition
\begin{equation}
(x(0), y(0), z(0) ) = {\bf r}_0,    \label{ABC-ini-2}
\end{equation}
where $ \varepsilon(t)$ is a normally random noise with a small standard deviation (at the order of magnitude $10^{-10}$), and ${\bf r}_0$ is the starting point, respectively.  We found that, even for the {\em fixed} values of $A=1, B=0.7, C=0.43$ and the {\em exact} starting position ${\bf r}_0 = (0,0,0)$, the statistics of the corresponding trajectory are unstable, i.e., rather sensitive to the normally random noise $ \varepsilon(t)$. In this case, the trajectory is ultra-chaotic, but there exists {\em no} sensitivity of statistics to parameters, because the starting position and {\em all} physical parameters have exactly the same values. This further indicates the essential differences between ultra-chaos and sensitivity of statistics to parameters.      

\subsection{Possible relationship between ultra-chaos and Poincar{\'e} section}

\begin{figure}
    \begin{center}
            \includegraphics[width=2.55in]{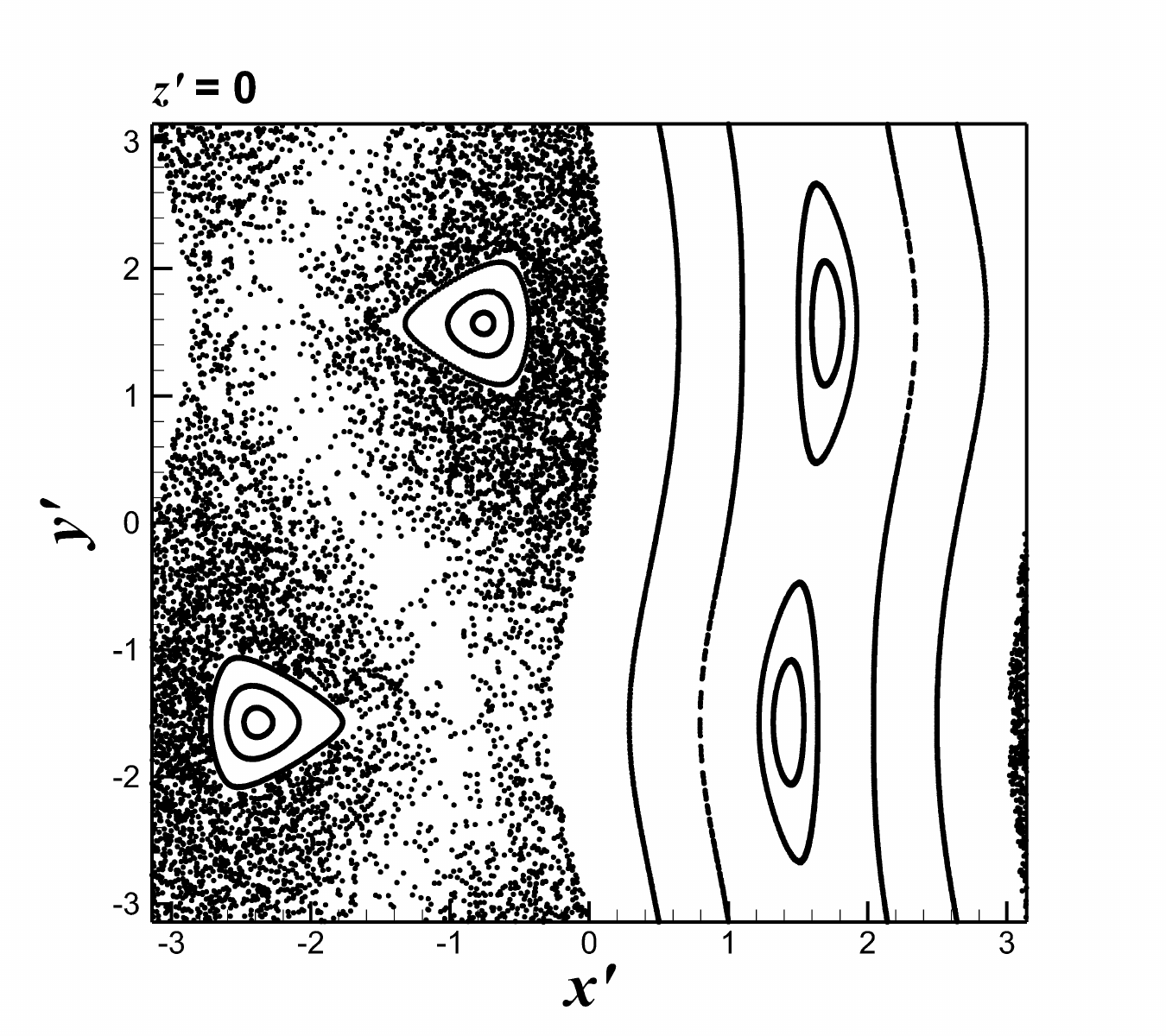}
    \end{center}
     \caption{Poincar{\'e} section at $z'=0$ for several normalized trajectories $(x'(t),y'(t),z'(t))$ in $t\in[0,10000]$ of fluid particles starting from different points ${\bf r}_{0}=(x(0),y(0),z(0))$ (as listed in table~\ref{PS_Ini}) obtained by means of CNS for ABC flow (\ref{ABC}) in the case of $A=1$, $B=0.7$ and $C=0.43$.}    \label{PS}
\end{figure}

Following \citet{dombre1986chaotic}, we obtain the Poincar{\'e} section of the ABC flow (\ref{ABC}) for $A=1$, $B=0.7$ and $C=0.43$, as shown in figure~\ref{PS}. Here, we applied CNS to obtain the trajectories $(x(t),y(t),z(t))$ of fluid particles in  $t\in[0,10000]$, starting from several selected points (listed in table~\ref{PS_Ini}).  For each trajectory, we have a point $(x',y')$ when $z'(t) = 2 n \pi$ for an arbitrary integer $n$, corresponding to a point $(x^{*},y^{*})$ in the square domain 
\[ x^{*} \in[-\pi, +\pi), \hspace{0.5cm}  y^{*}\in[-\pi, +\pi), \] 
by means of the periodic condition in $x'$ and $y'$ directions, say,
\[ x' = x^{*} + 2 m \pi, \hspace{0.5cm}  y' = y^{*} + 2 k  \pi,  \]
where $m$ and $k$ are integers.  
 The set of all these points $(x^{*},y^{*})$ gives the Poincar{\'e} section of the ABC flow (\ref{ABC}), as shown in figure~\ref{PS}. For details, please refer to \citet{dombre1986chaotic}.

As shown in figure~\ref{PS},  there exist elliptic islands (or KAM tori) and a chaotic sea in the Poincar{\'e} section. By convention, it is widely believed that points in an elliptic island correspond to quasi-periodic orbits or weakly chaotic orbits, but points in a chaotic sea correspond to strongly chaotic orbits, respectively \citep{PhysRevE_61_3777,  Skokos_2001,  LUKESGERAKOPOULOS20081907}. Interestingly, the Poincar{\'e} section (as shown in figure~\ref{PS}) is rather similar to figure~\ref{r0_1}(a).  So, it is reasonable for particles starting from the elliptic islands (or KAM tori) to represent a kind of normal-chaotic property, because their maximum Lyapunov exponents $8.5\times 10^{-5} \leq \lambda_{max} \leq 1.3\times 10^{-2}$ (listed in table~\ref{Lye}) indeed correspond to a weak chaos.  This numerical {\em fact} reveals the following relationship: the normal-chaotic (starting) points  (at $z=0$) of the ABC flow correspond to the elliptic islands (or KAM tori) in the Poincar{\'e} section, but the ultra-chaotic ones invariably correspond to the chaotic sea.     According to our computations, this kind of relationship is  true for almost all fluid particles in the ABC flow.  Besides,  this numerical experiment  also supports our conclusion that an ultra-chaos is a higher disorder than a normal-chaos.  Thus, the classification of chaos into normal-chaos and ultra-chaos provides a {\em new} explanation of elliptic islands (or KAM tori) and chaotic sea in Poincar{\'e} section of a dynamic system.    

Note that Poincar{\'e} section has a close relationship with KAM theory that is valid for an {\em integrable} Hamiltonian system only. However, the classification of chaos into ultra-chaos and normal-chaos is generally valid for {\em all} dynamic systems, even if they are {\em not} Hamiltonian, or {\em not} integrable. Therefore,  this classification has a more general meaning, given that ultra-chaos reveals higher disorder than normal-chaos. An example of such higher disorder related to statistical sensitivity to small disturbances has been recently reported:  small disturbances can lead to large-scale deviations of simulations of a turbulent flow {\em not only} in spatiotemporal trajectories {\em but also} in statistics, even leading to different types of flow \citep{Qin2022JFM}.

\begin{table}
\tabcolsep 0pt
\vspace*{-2pt}
\begin{center}
\begin{tabular}{ccc}
\hspace{1.0cm} Starting point \hspace{1.0cm} & \hspace{1.0cm} $x(0)$\hspace{1.0cm}  &\hspace{1.0cm}  $y(0)$\hspace{1.0cm}    \\
\\
~~No.1 & $0$ & $0$  \\
~~No.2 & $-0.1$ & $0$  \\
~~No.3 & $-0.2$ & $0$  \\
~~No.4 & $-0.3$ & $0$  \\
~~No.5 & $-0.4$ & $0$  \\
~~No.6 & $-0.5$ & $0$  \\
~~No.7 & $-0.6$ & $0$  \\
~~No.8 & $-0.7$ & $0$  \\
~~No.9 & $-0.8$ & $0$  \\
~~No.10 & $-0.9$ & $0$  \\
~~No.11 & $-1.0$ & $0$  \\
~~No.12 & $-1.5$ & $0$  \\
~~No.13 & $0.5$ & $0$  \\
~~No.14 & $1.0$ & $0$  \\
~~No.15 & $1.5$ & $1.5$  \\
~~No.16 & $1.6$ & $1.6$  \\
~~No.17 & $-1.8$ & $-1.5$  \\
~~No.18 & $-2.1$ & $-1.5$  \\
~~No.19 & $-2.3$ & $-1.5$  \\
~~No.20 & $-0.8$ & $1.5$  \\
~~No.21 & $-1$ & $1.5$  \\
~~No.22 & $-1.3$ & $1.5$  \\
\end{tabular}
\end{center}
\caption{Positions ${\bf r}_{0}=(x(0),y(0),0)$ of the starting particles, chosen for the Poincar{\'e} section shown in figure~\ref{PS}.}    \label{PS_Ini}
\end{table}

\subsection{Possible relationship between ultra-chaos and ergodicity/non-ergodicity}

According to our numerical experiments mentioned above, statistics are stable for a normal-chaotic motion of fluid particle in the ABC flow, but unstable for an ultra-chaotic motion of fluid particle in the {\em same} ABC flow. However, it is an {\em open} question whether or not a normal-chaos should correspond to ergodicity and an ultra-chaos to non-ergodicity, because it is rather difficult in practice to prove ergodicity or non-ergodicity of a dynamic system.

According to \cite{Birkhoff1931PNAS} and \cite{vonNeumann1932PNAS},  time averages can be set equal to phase averages, provided the system is ergodic, i.e. {\em metrically transitive} \citep{Moore2015PNAS}. However, it is difficult to prove conclusively that a system is  metrically transitive. In fact, \cite{Birkhoff1931PNAS} and \cite{vonNeumann1932PNAS} did not actually solve the problem of equating time average and phase averages but instead reduced it to an equally difficult problem of proving metric transitivity, as pointed out by \cite{Moore2015PNAS}. For example, in the case of $A=1, B=0.7$ and $C=0.43$, our computations clearly indicate that a fluid particle starting from ${\bf r}_{0,n}=(0,-0.1,0)$ has a normal-chaotic trajectory, but another fluid particle starting from ${\bf r}_{0,u}=(-0.1, 0.1,0)$ has an ultra-chaotic trajectory, as shown in the movie. How can we prove that, under the same velocity field ${\bf u}_{ABC}$ of the ABC flow, the trajectory of the former particle is metrically transitive (i.e. ergodic) but the latter is not (i.e. non-ergodic)? From a practical viewpoint, it is much easier for us to conclude that the former trajectory is normal-chaotic whereas the latter trajectory is ultra-chaos by investigating the stability of its statistical properties via time average (or ensemble average) than to prove (or disprove) its metric transitivity for ergodicity, since either time or ensemble average is quite easy. Therefore, from practical point of view, ultra-chaos is a more useful concept than non-ergodicity!     

It would be very convenient if one could theoretically prove (or disprove) that {\em every} ergodic system corresponds to a normal-chaos and {\em every} non-ergodic system to an ultra-chaos, respectively.  If so, the classification of normal-chaos and ultra-chaos might provide us with a simple and  practical way to reveal ergodic properties of various types of dynamic systems, given that it is much easier to check stability of statistics through spatio-temporal average (or ensemble average) than to prove (or disprove) metric transitivity.            

\section{Possible relationship between ultra-chaos and turbulence}

The velocity $\mathbf{u}_{ABC}$ of the Arnold-Beltrami-Childress (ABC) flow (\ref{ABC-0})
was first discovered by \citet{arnold1965} as a class of steady-state solutions of the Euler equations, and moreover, with external force per unit mass, it also satisfies the  Navier-Stokes momentum and continuity equations
\begin{equation}
\frac{\partial\mathbf{u}}{\partial t}+(\mathbf{u}\cdot\nabla)\hspace{0.3mm}\mathbf{u}=-\nabla p +\frac{1}{Re}\hspace{0.2mm}\Delta\hspace{0.3mm}\mathbf{u}+\mathbf{f},    \label{NS-1}
\end{equation}
\begin{equation}
\nabla\cdot \mathbf{u}=0,    \label{NS-2}
\end{equation}
where $t\geq0$ denotes the time, $\nabla$ is the Hamilton operator, $\Delta$ is the Laplace operator, $Re$ is the Reynolds number, $p$ denotes the pressure and 
\begin{equation}
\mathbf{f}=\frac{\mathbf{u}_{ABC}}{Re} \label{external-force}
\end{equation} 
is the given external force per unit mass, with the periodic boundary conditions at $x =\pm\,\pi$, $y =\pm\,\pi$, and $z =\pm\,\pi$.

\begin{figure}
    \begin{center}
            \subfigure[]{\includegraphics[width=1.7in]{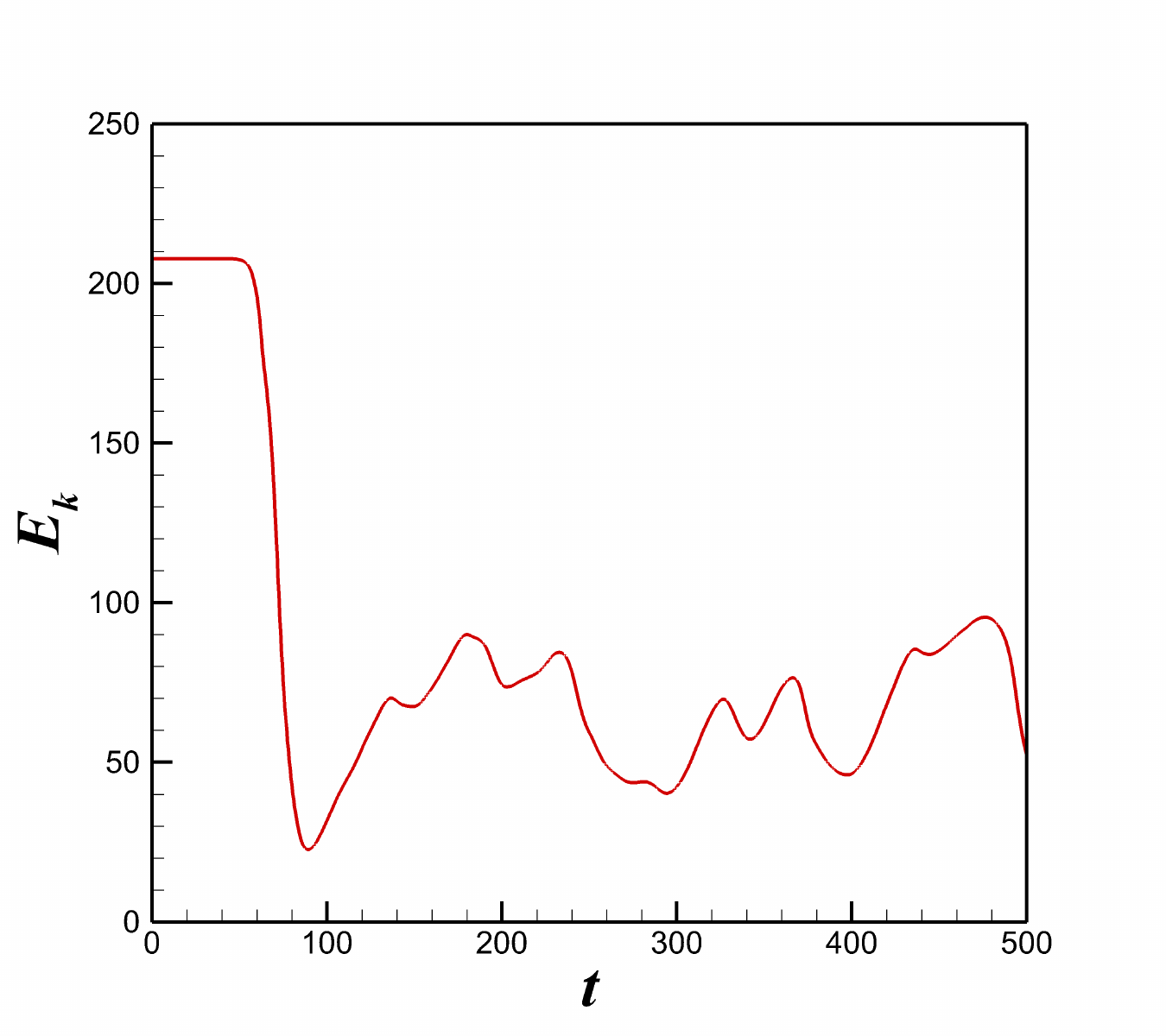}}
            \subfigure[]{\includegraphics[width=1.7in]{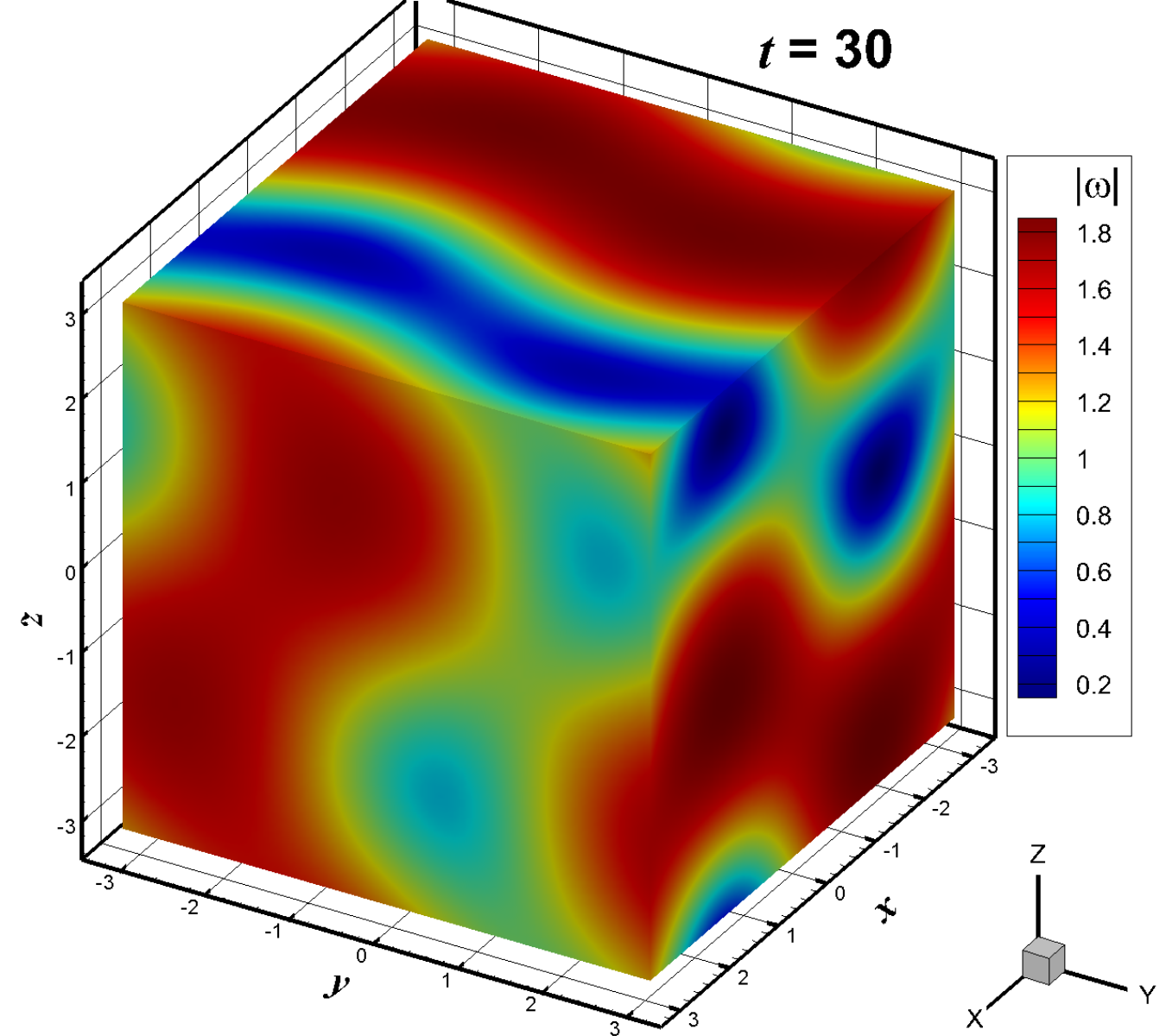}}
            \subfigure[]{\includegraphics[width=1.7in]{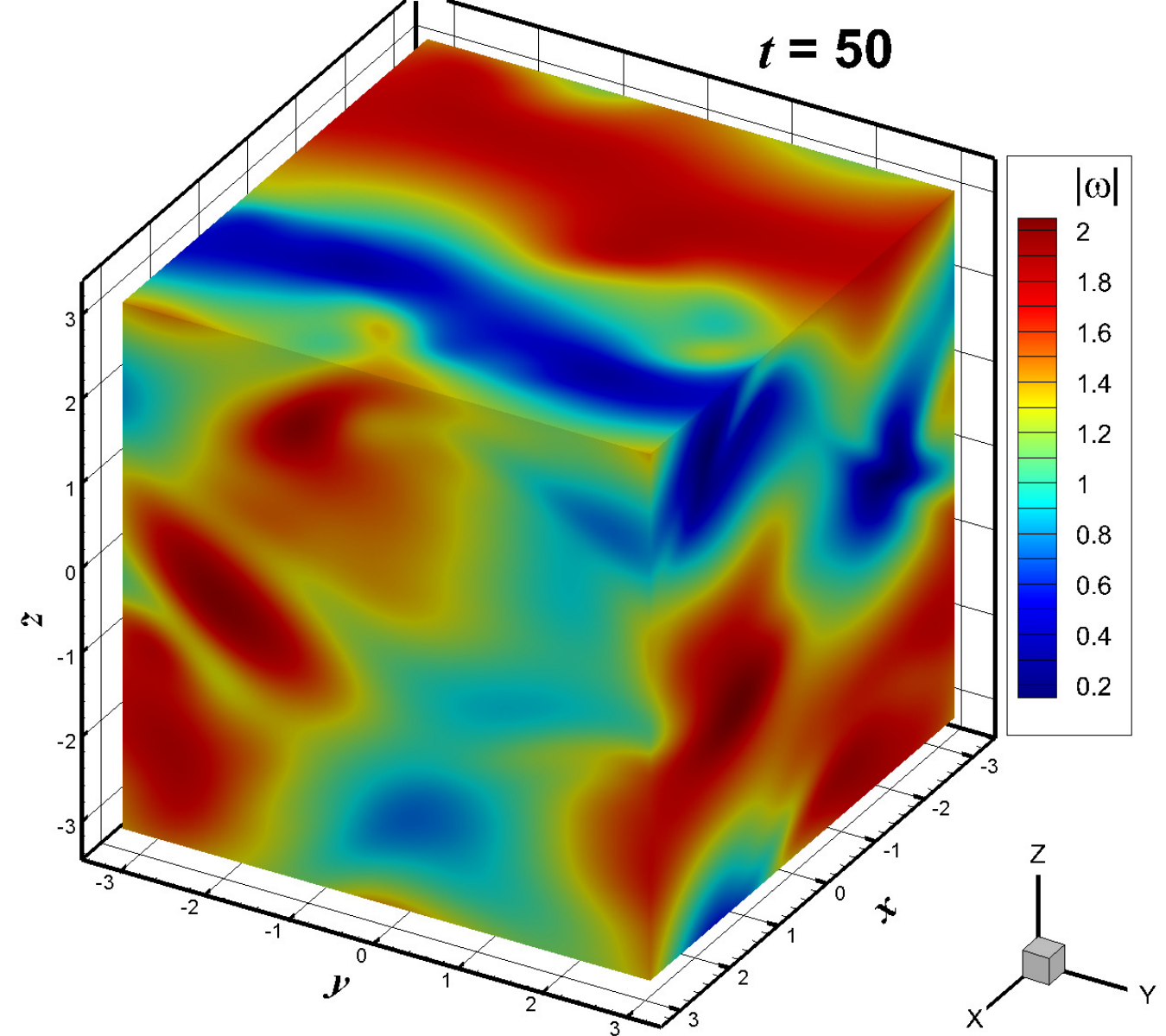}}
            \subfigure[]{\includegraphics[width=1.7in]{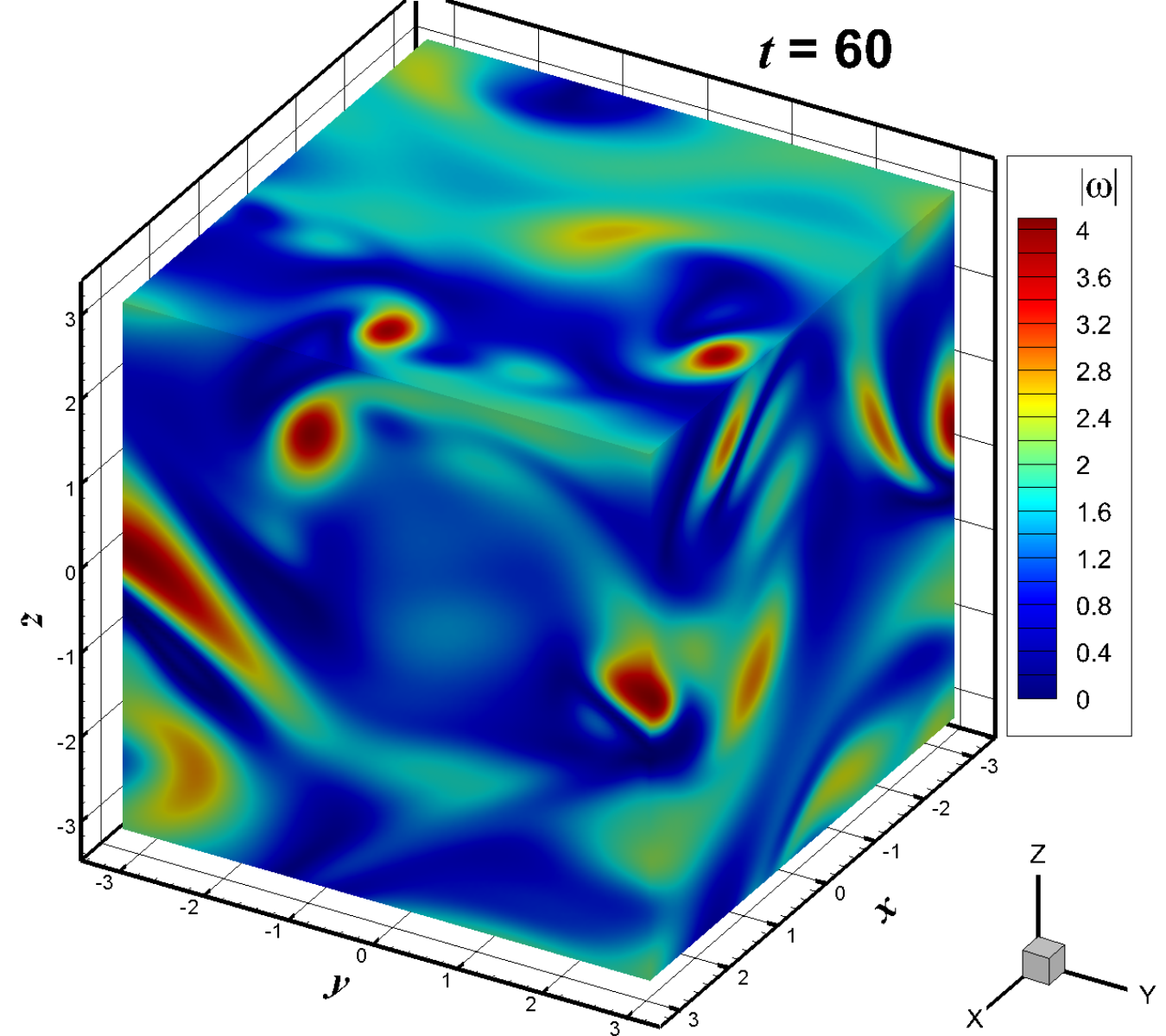}}
            \subfigure[]{\includegraphics[width=1.7in]{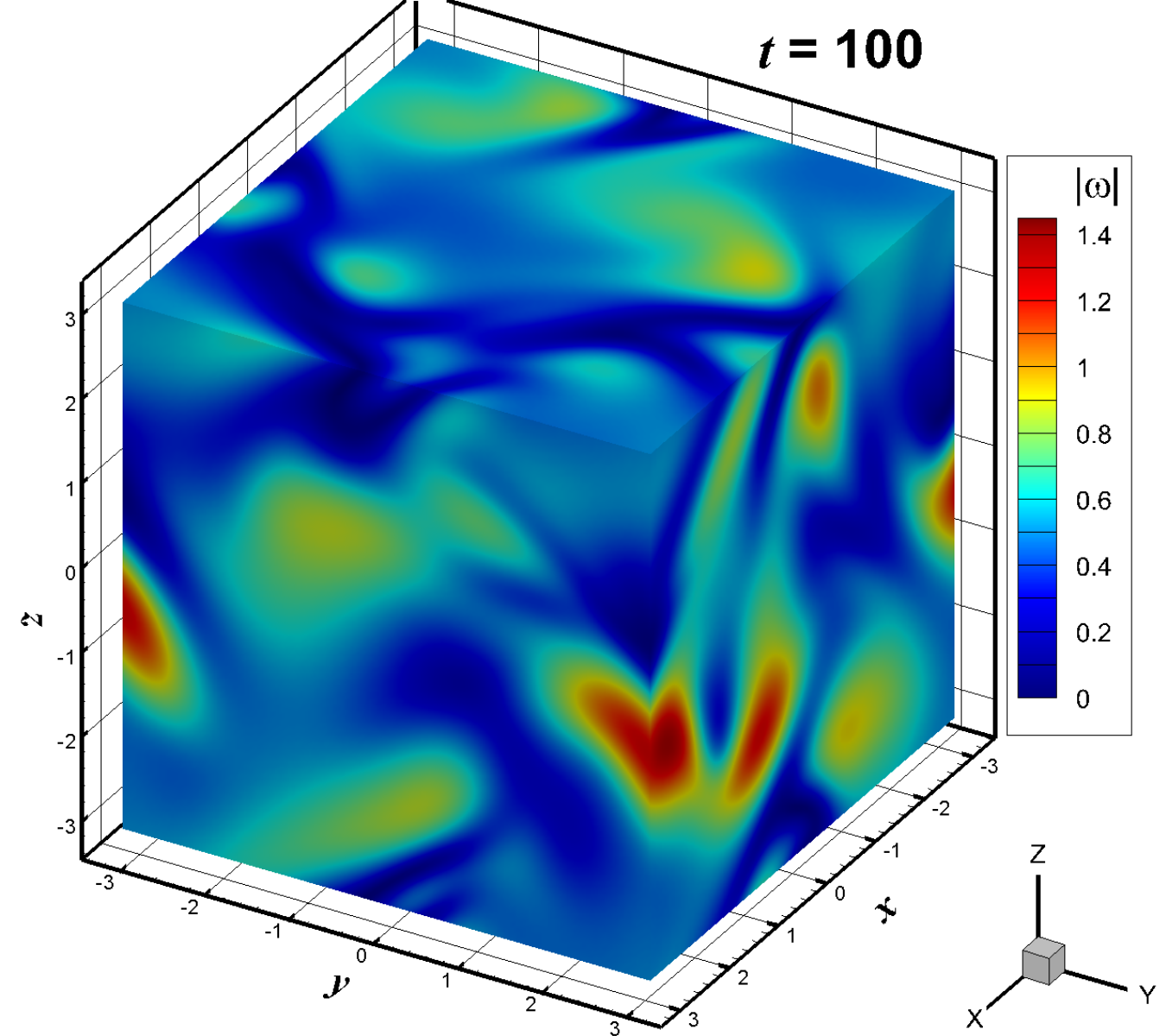}}
            \subfigure[]{\includegraphics[width=1.7in]{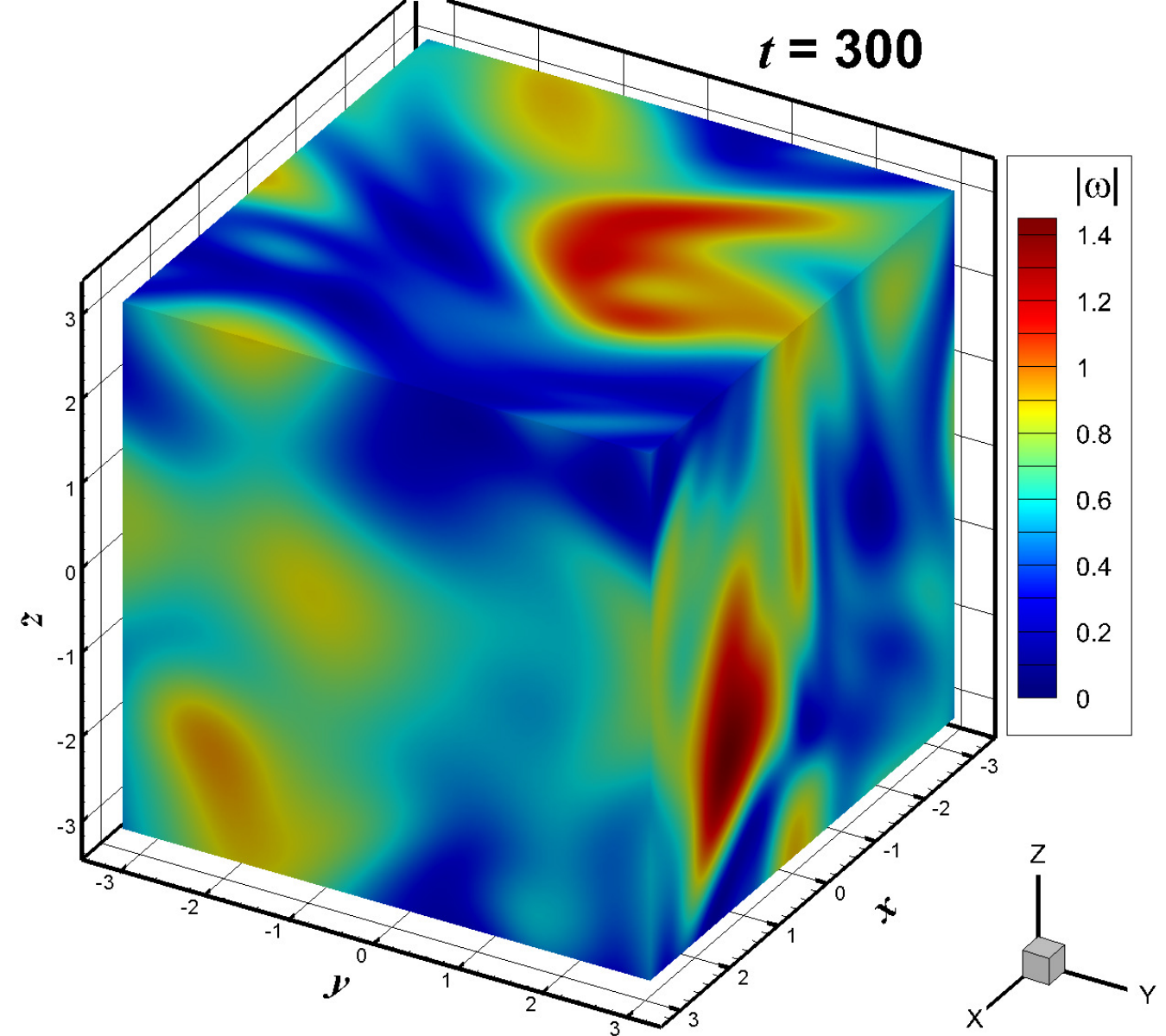}}
        \caption{(a) Total kinetic energy; (b)-(f) Modulus $\left|\bm{\omega}\right|$ of instantaneous vorticity field at $t=30$, $t=50$, $t=60$, $t=100$, or $t=300$, respectively, governed by the NS equations (\ref{NS-1}) and (\ref{NS-2}) with $R_{e}=50$ using the ABC flow (\ref{ABC-0}) (for $A=1$, $B=0.7$ and $C=0.43$) under a small disturbance at the order of magnitude $10^{-3}$ as the initial solution.}    \label{C043-results}
    \end{center}
\end{figure}

Many papers \citep{dombre1986chaotic, Mezic2002PoF, Podvigina2006PD} have been published in this field.  For example, \cite{Podvigina2006PD}  analyzed the bifurcation of the ABC flow and reported the supercritical Hopf bifurcation and route to chaos through tori doubling.    
Without loss of generality, let us consider here the ABC flow in the case of  $A=1$, $B=0.7$ and $0 \leq C \leq 0.43$. Unlike other researchers \citep{dombre1986chaotic, Mezic2002PoF, Podvigina2006PD}, here we mainly focus on the ultra-chaotic motion of fluid particles.
As reported by \citet{podvigina1994non}, the Reynolds number $Re=50$ corresponds to a turbulent flow if the initial velocity field $\mathbf{u}_{ABC}$ experiences small disturbances of the order of magnitude $10^{-3}$. This kind of turbulent flow is solved numerically in $t\in[0,500]$: the spatial domain $[-\,\pi, +\,\pi)^3$ is discretized by a uniform mesh with $128^3$ points for the spatial Fourier expansion, where the maximum grid spacing is less than the minimum Kolmogorov scale \citep{pope2001turbulent}, and the $3/2$ rule for dealiasing \citep{pope2001turbulent} is used, with the time step $\triangle t=10^{-3}$.

First, let us use the unstable ABC flow for $A=1.0$, $B=0.7$ and $C=0.43$ as the initial condition of the NS equations (\ref{NS-1}) and (\ref{NS-2}) (with the external force per unit mass $\mathbf{f}=\mathbf{u}_{ABC}\hspace{0.3mm}/\hspace{0.2mm}Re$ as mentioned above), under small disturbances of initial velocity at the order of magnitude $10^{-3}$. It is found that, the flow is initially rather similar to the ABC flow at times that are too short for the tiny velocity disturbances to transfer to the macro-level; in this situation, about 49\% starting fluid particles are ultra-chaotic, according to table~\ref{C_ultra-T}. The transition from laminar flow to turbulence occurs approximately at $t \approx 50.0 = T_{tran}$, as shown in figure~\ref{C043-results}, where $T_{tran}$ denotes the time of the transition occurrence.  Given the velocity field $\mathbf{u}$ of equations (\ref{NS-1}) and (\ref{NS-2}), we can investigate the chaotic property of trajectory (i.e. whether it is ultra-chaotic or normal-chaotic) of a fluid particle starting from $\mathbf{r}_{0}$ in a similar way as mentioned in \S~2.  
When $t=50$, we randomly choose 10000 starting fluid particles in $-\pi\leq x,y,z < +\,\pi$ and find that {\em all} trajectories of the fluid particles starting from them are ultra-chaotic. This numerical {\em experiment} strongly suggests that a necessary condition of turbulence is that almost {\em all} fluid particles should move along ultra-chaotic trajectories.

Similarly, let us consider the stable ABC flow in the case of $A=1$, $B=0.7$ and $C=0$. Using the ABC flow $\mathbf{u}_{ABC}$ subject to small disturbances at the order of magnitude $10^{-3}$ as the initial condition, we numerically solve the NS and continuity equations (\ref{NS-1}) and (\ref{NS-2}) in the time interval $t\in[0,2000]$ and investigate the chaotic property of trajectories of the 10000 randomly chosen fluid particles.  It is  found that, when $C=0$, transition from the laminar flow to turbulence {\em never} occurs, and moreover, there exists {\em no} ultra-chaotic motion for {\em all} of these fluid particles throughout $t\in[0,2000]$. This  further confirms our suggestions  that ultra-chaotic trajectories of fluid particles should have some relationships with turbulence.

\begin{figure}
    \begin{center}
            \includegraphics[width=2.55in]{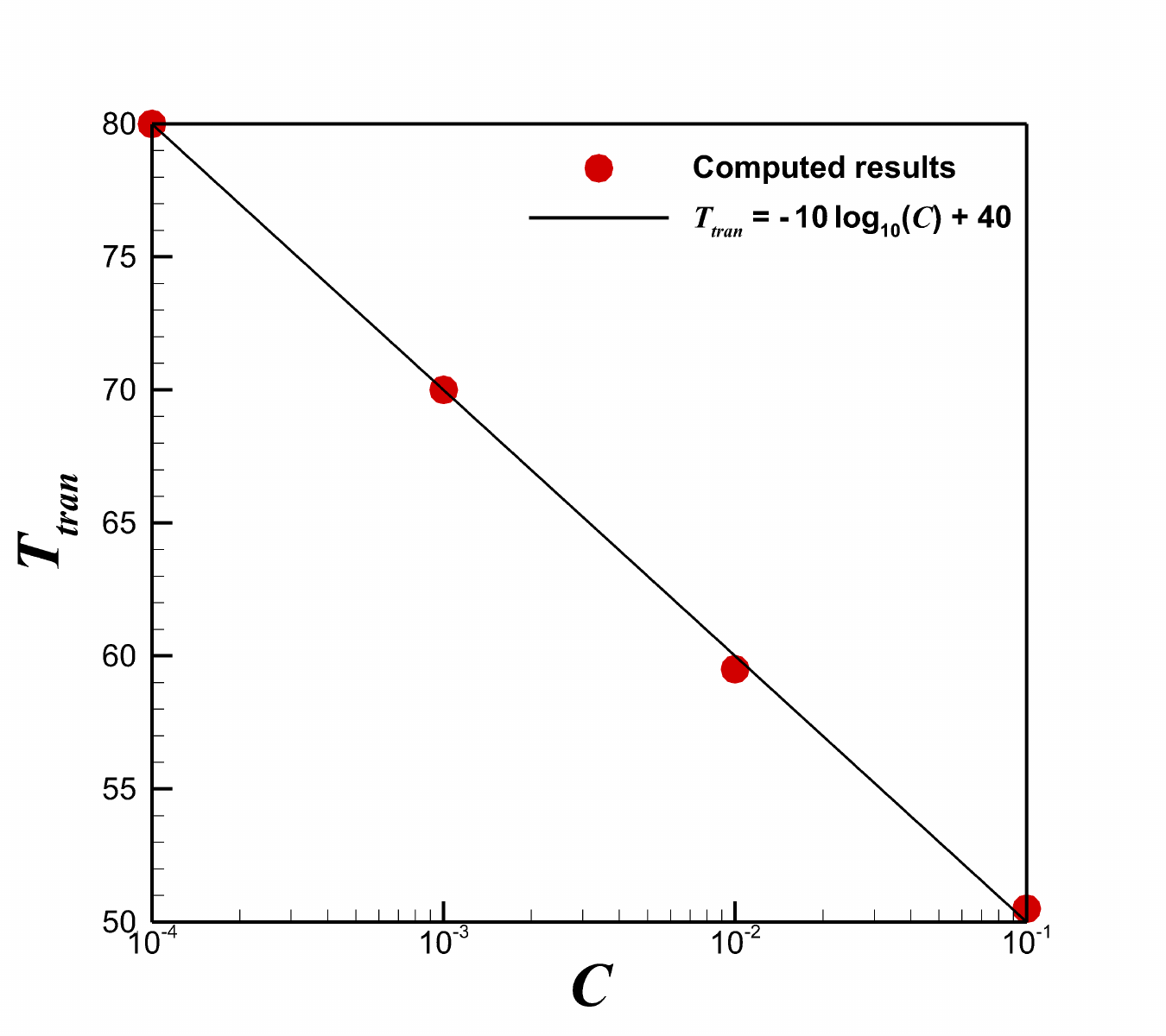}
        \caption{Relationship between $C$ and $T_{tran}$ obtained for ABC flow with $A=1$, $B=0.7$ and $0<C\leq0.1$ subject to the ABC flow (\ref{ABC-0}) plus a small disturbance at order of magnitude $10^{-3}$ as the initial condition used when solving the NS and continuity equations (\ref{NS-1}) and (\ref{NS-2}), where $T_{tran}$ denotes the time of  transition occurrence.}    \label{C_Tran}
    \end{center}
\end{figure}

\begin{table}
\tabcolsep 0pt
\vspace*{-2pt}
\begin{center}
\begin{tabular}{cc}
\hspace{1.0cm} $C$\hspace{1.0cm} & \hspace{1.0cm} $T_{tran}$\hspace{1.0cm}  \\ \\
~~$0.1$ & $50.5$~~ \\
~~$0.01$ & $59.5$~~ \\
~~$0.001$ & $70.0$~~ \\
~~$0.0001$ & $80.0$~~ \\
~~$0.0$ & $-$~~ \\
\end{tabular}
\end{center}
\caption{Transition occurrence time $T_{tran}$ for different values of $C$, gained by numerically solving the NS equations  (\ref{NS-1}) and (\ref{NS-2})  using the ABC flow (\ref{ABC-0}) (for $A=1$, $B=0.7$ and $0\leq C\leq0.1$) as the initial condition plus a small disturbance of velocity field at the order of magnitude $10^{-3}$. }    \label{C_Tran-T}
\end{table}

In addition, let us further consider the ABC flows for $A=1$, $B=0.7$ and different values of $C$, and for each case we use the Monte-Carlo method to randomly choose 10000 starting points in $-\,\pi\leq x,y,z < +\,\pi$.  It is found that the transition time $T_{tran}$ when the flow alters from laminar flow to turbulence increases as $C$ decreases from $0.1$ to $0.0001$, as shown in table~\ref{C_Tran-T}.  We found that, when $0 < C\leq0.1$, there exists a linear relationship
\begin{equation}
T_{tran}\approx-10\log_{10}(C)+40,    \label{relation_C_Tran}
\end{equation}
as illustrated in figure~\ref{C_Tran}, indicating that $T_{tran} \to +\infty$ as $C\to 0$. Hence, indeed the transition from laminar flow to turbulence should  never occur when $C=0$, which agrees with our numerical simulation in the case of $C=0$ mentioned above. In all cases of $A=1$, $B=0.7$ and $0 < C \leq 0.43$ under consideration, it is found that all trajectories of the fluid particles starting from the randomly chosen 10000 fluid particles are ultra-chaotic after the flow becomes fully turbulent.  Besides, the smaller the number of ultra-chaotic particles at the beginning (corresponding to an unstable ABC flow with a smaller value of  $C$), the longer the transition time of $T_{tran}$. In other words, more time is needed for all fluid particles to become ultra-chaotic at $t =T_{tran}$. The foregoing again suggests that a necessary condition of the occurrence of transition from laminar flow to turbulence is that  {\em nearly all fluid particles should move along ultra-chaotic trajectories}.

Here, regarding the ultra-chaotic motion as a new property of fluid particle, we simply report some results of numerical experiments governed by equations (\ref{NS-1}) and (\ref{NS-2}) subject to periodic boundary conditions. Frankly speaking, it is not yet understood why almost all fluid particles should move along ultra-chaotic trajectories when the flow becomes fully turbulent, what happens to trajectory property of fluid particles when the transition from laminar flow to turbulence occurs, and so on. We firmly believe that the concept of ultra-chaos could enable us to gain new insight into viscous flows. We intend to pursue this in future work.           

It should be emphasized that the transition from laminar flow to turbulence is a key problem in fluid mechanics. Hopefully, ultra-chaos as a new concept could provide us a completely {\em new} viewpoint to investigate and understand the mechanism of the transition to turbulence.

\section{Concluding remarks and discussions}

Due to the butterfly-effect \citep{lorenz1963deterministic}, numerical noise (due to truncation error and round-off error) enlarges exponentially so that  a computer-generated simulation of chaotic systems quickly becomes a mixture of a  ``true'' physical solution $s$ and a ``false'' numerical noise $\varepsilon$, which are mostly at the {\em same} order of magnitude.  In practice, statistics of chaotic systems are usually calculated using such a mixture, i.e. $s+\varepsilon$, because there is no way to separate out the ``true'' physical solution $s$. In fact, the statistics is based on the {\em hypothesis} 
\[ \big< s+\varepsilon \big> =  \big< s \big>,   \]  
where $\big< \big>$ is a statistical operator. Here, the numerical noise $\varepsilon$ is in fact equivalent to a kind of small disturbances. Unfortunately, there exists {\em no} theoretical proof that the above {\em hypothesis} is {\em always} true in general.            

By means of CNS \citep{Liao2009, Liao2013, Liao2014, Liao2017-CNS-review}, one can gain reliable/convergent numerical simulations of chaotic systems over a long enough interval of time \citep{LIAO2014On, Li2017More, lin2017origin, hu2020risks, qin2020influence, Liao2022AAMM, Qin2022JFM,  Yang2023CFS}, during which the ``false'' numerical noise $\varepsilon$ is much smaller than the ``true'' physical solution $s$, say, $|\varepsilon| \ll |s|$, thus the influence of the numerical noise $\varepsilon$ is negligible compared to the ``true'' physical solution $s$. Hence, CNS can provide us, for the first time, with a nearly ``clean'' computer-generated simulation of chaos in an interval of time long enough for statistics, which can be used as a benchmark solution since it is very close to the ``true'' physical solution $s$. Therefore, CNS provides us with a useful tool by which to study accurately the influence of numerical noise on statistics of different types of chaotic systems \citep{hu2020risks, qin2020influence, Liao2022AAMM, Qin2022JFM, Yang2023CFS}.        

Using CNS, it has been observed that the hypothesis $\big< s+\varepsilon \big> =  \big< s \big>$  holds for many chaotic systems, whose statistics are stable to small disturbances. However, it has been found that statistics of some chaotic systems are indeed rather sensitive to small disturbances including artificial numerical noise \citep{hu2020risks, qin2020influence, Qin2022JFM, Yang2023CFS}, say, $\big< s+\varepsilon \big> \neq  \big< s \big>$. \cite{Liao2022AAMM} termed the former ``normal-chaos'' and the latter ``ultra-chaos'', respectively.  The stability of trajectory and statistics of different types of dynamic systems is listed in table~\ref{disorder}, which clearly indicates that an ultra-chaos is a higher disorder than a normal-chaos.

It is well known that the steady-state ABC flow, which satisfies the steady NS equations with a proper external force, has chaotic properties from a Lagrangian viewpoint.   Naturally,  it is worth investigating whether ultra-chaos exists in ABC flow, whether relationships hold between ultra-chaos and turbulence, and so on.  
In this paper, we illustrate that trajectories of many fluid particles in the steady-state ABC flow (\ref{ABC-0}) are ultra-chaotic, in that their statistical properties are rather sensitive to tiny disturbances. Obviously, this kind of ultra-chaotic motion of fluid particles represents a higher disorder than normal-chaotic ones. We found that these two kinds of totally different chaos coexist widely and simultaneously in the steady-state ABC flow. 

Besides, we discuss the difference between ultra-chaos and the high sensitivity of statistics to parameters, and also the relationships between ultra-chaos and Poincar\'{e} section, ultra-chaos and ergodicity, and so on. 
Unlike the high sensitivity of statistics to parameters, ultra-chaos focuses on the stability of statistics to very small disturbances. We have illustrated that certain model equations (such as the Lorenz-84 climate model with seasonal forcing) exhibit high sensitivity of statistics to parameters, even though all the statistics remain stable so that all the simulations involve either non-chaos or normal-chaos.   

Furthermore, we discuss the possible relationship between the Poincar\'{e} section and normal/ultra-chaos. It is found that the normal-chaotic (starting) points (at $z=0$) of the ABC flow correspond to elliptic islands (or KAM tori), but the ultra-chaotic ones correspond to the chaotic sea of the Poincar\'{e} section. It should be emphasized that, considering the fact that the Poincar{\'e} section has a close relationship with the KAM theory that is usually suitable for an {\em integrable} Hamiltonian system only, the new classification of chaos into ultra-chaos and normal-chaos is more general because of its validity for {\em all} dynamic systems, even if they are {\em not} Hamiltonian, and/or {\em not} integrable. Therefore, this new classification should have more general meaning.

In addition, we discuss possible relationships between ultra-chaos and non-ergodicity. According to \cite{Birkhoff1931PNAS} and \cite{vonNeumann1932PNAS}, time averages can be set equal to phase averages, provided the system is ergodic (i.e., metrically transitive) \citep{Moore2015PNAS}. However, it is very difficult to seriously prove that a system is {\em metrically transitive}.  It is an open question whether or not a normal-chaos corresponds to ergodicity and an ultra-chaos corresponds to non-ergodicity, respectively.  Hopefully the classification of chaos into  normal-chaos and ultra-chaos could provide us a simple and practical way to investigate ergodic property of various types of dynamic systems.

In order to study the possible relationships between ultra-chaos and turbulence, we numerically solve the NS and continuity equations (\ref{NS-1}) and (\ref{NS-2}) when $Re = 50$ using the ABC flow (for $A=1, B =0.7$ and various values of $C$) plus a small disturbance as the initial condition. It is found that the trajectories of nearly all fluid particles become ultra-chaotic after the transition from laminar to turbulent flow occurs. Our numerical {\em facts} strongly suggest that turbulence is likely to be related to ultra-chaotic trajectories of fluid particles, although the detailed mechanisms are presently unknown.
Considering that the chaotic property of the ABC flow is essential for the development of turbulence \citep{dombre1986chaotic, galloway1987note, podvigina1994non}, we anticipate that the concept of ultra-chaos \citep{Liao2022AAMM} could lead to a better understanding of fluid chaos, Poincar\'{e} section, ergodicity/non-ergodicity, turbulence, and their inter-relationships.

\backsection[Supplementary movie]{A supplementary movie is available at ...}

\backsection[Acknowledgements]{Thanks to the anonymous reviewers for their valuable suggestions and constructive comments.}

\backsection[Funding]{This work is partly supported by National Natural Science Foundation of China (No. 12272230), Shanghai Pilot Program for Basic Research - Shanghai Jiao Tong University (No. 21TQ1400202) and the Fundamental Research Funds for the Central Universities.}

\backsection[Declaration of Interests]{The authors report no conflict of interest.}

\backsection[Data availability statement]{The data that support the findings of this study are available on request from the corresponding author.}

\backsection[Author ORCID]{Shijie Qin, https://orcid.org/0000-0002-0809-1766; Shijun Liao, https://orcid.org/0000-0002-2372-9502}

\backsection[Author contributions]{Liao conceived and designed the analysis and related concepts. Qin performed the computations. Both wrote the manuscript.}

\bibliographystyle{jfm}
\bibliography{Ultrachaos-ABC}

\end{document}